\documentclass[11pt, hyper]{JHEP3}

\usepackage{fancyhdr}
\usepackage{graphicx}
\usepackage{amsmath}
\usepackage{lootex}
\usepackage{braket}
\usepackage{youngtab}
\usepackage{epsfig}

\def\mbf#1{\mathchoice{\hbox{\boldmath $\displaystyle #1$}}
        {\hbox{\boldmath $\textstyle #1$}}
        {\hbox{\boldmath $\scriptstyle #1$}}
        {\hbox{\boldmath $\scriptscriptstyle #1$}}}

\preprint{QMUL-PH-09-17}

\title{Direct Mediation and Metastable Supersymmetry Breaking for SO(10)}

\author{Daniel Koschade\footnote{d.koschade@qmul.ac.uk}, Moritz McGarrie\footnote{m.mcgarrie@qmul.ac.uk}, and Steven Thomas\footnote{s.thomas@qmul.ac.uk} \\ \\

Queen Mary University of London\\
Center for Research in String Theory \\
Department of Physics, \\
Mile End Road, London, E1 4NS, UK.
}

\abstract{We examine a metastable $\mathcal{N}=1$ macroscopic $SO(N)$ SQCD model of Intriligator, Seiberg and Shih (ISS). 
We introduce various baryon and meson deformations, including multitrace operators and explore embedding an $SO(10)$ parent of the standard model into two weakly gauged flavour sectors. Direct fundamental messengers and the symmetric pseudo moduli messenger mediate SUSY breaking to the MSSM. Gaugino and sfermion masses are computed and compared for each deformation type.  We also explore \emph{Uplifting}: reducing the rank of the magnetic quark matrix of the ISS model and find an additional fundamental messenger.}


\begin{document}

\section{Introduction}

The LHC is here.  Its arrival offers the opportunity to verify supersymmetry and any signatures of a gauge unifying ``parent" of the standard model.
 ``ISS'' type direct gauge mediation of supersymmetry breaking, in a dynamically SUSY broken, metastable vacuum (MSB), is an approach to this. 
It is succinct, perturbative and calculable \cite{intriligator2006dsb}.  Schematically, supersymmetry is broken in a hidden sector and 
  the hidden sector dynamics generate messenger fields when a global symmetry is weakly gauged and associated with the visible standard model.  These messengers bring about 
  the MSSM soft terms by running in one and two loop diagrams.  The power of applying Seiberg duality in the hidden sector is that the model is 
  perturbative in either the electric or, its dual, magnetic description. In particular one can have a UV free electric description and an 
  IR free effective description in which supersymmetry breaking can be explored perturbatively.  The supersymmetry breaking vacuum need only
   be metastable, allowing for R-symmetry to be broken and gaugino masses to be generated.

The hidden sector is a SUSY QCD Seiberg dual model.  These have been found for $SU(N)$, $SO(N)$ and $Sp(N)$ gauge groups.  
Initially $SU(N)$ models were explored in which the messengers from the hidden to visible sector are in standard model representations \cite{Kitano:2006xg,Abel:2007uq,Amariti:2006vk}. 
 Later, models in which the messengers formed complete representations of a $SU(5)$ were implemented \cite{abel_direct_2008}.  This paper extends 
 the ``Dictionary'' of possible MSB models by implementing an $SO(N)$ model in which the messengers are in complete representations of $SO(10)$.

$SO(10)$ is a grand unified theory (GUT) that can be broken to $SU(5)$  or ``left-right symmetric''  (SUSYLR) models, both of which have a low
 energy standard model limit.  SUSYLR has distinctive and predictive signatures; it can generate neutrino masses in a natural way and 
 reinstate left right symmetry of the leptonic sector at a suitable energy. In this paper we identify part of the hidden sector flavour symmetry 
 with $SO(10)$, thereby making the hidden sector dynamics compatible with visible sector GUT models based on $SO(10)$.
 
In section 2 we review the macroscopic model and the choice of embeddings of $SO(10)$ into a weakly gauged flavour group. 
 We examine the field content and identify the messenger fields which will generate soft terms for the MSSM.  In section 3 we look at
  how various messengers will affect gaugino masses and use this  to guide our analysis of the deformations of the ISS model. 
   Section 4 introduces some minimal R-symmetry breaking deformations and explore the outcomes of these deformations of the vacuum of 
   the theory.  Section 5 uses these deformations and a ``KOO'' \cite{Kitano:2006xg} deformation and calculates the contributions to the messenger and gaugino masses. Section 6 explores the non-perturbative potential for the ISS model and 
   determines the lifetime of the SUSY broken vacuum.  Section 7 explores the particle content of the messenger fields in terms of the 
   standard model gauge group and 1-loop beta function contributions are explored.  In section 8 we
    extend the deformation types of section 5 by adding multitrace deformations to the ISS model.  We then explore how these models 
	behave when the rank of SUSY breaking magnetic quark matrices are reduced. Section 9 discusses a possible extension of ISS models with dual pairs.  In section 10 we conclude. Appendix A reviews some aspects of computing soft terms by the methods of 
	``General Gauge Mediation'' \cite{Meade:2008wd} in the $SO(10)$ case and Appendix B, as an  aside,  explores the viability of a non-ISS minimum of
	  the KOO model.


\section{Seiberg dual pictures for SO(N) SQCD}



Seiberg duality is an electromagnetic duality at an interacting fixed point in which by 
interchanging electrically charged variables with ``magnetic'' fundamental variables, one may interchange
 different unphysical gauge groups and their strong/weak coupling, preserving the physical global symmetries 
 in both pictures.  This section will outline both sides of the duality. In particular we will choose the weakly 
 coupled side of the duality to explore supersymmetry breaking.  This section closely reviews  \cite{intriligator2006dsb} and sets
 our notation.
 

 \subsection{Microscopic theory}
We describe the electric side of the duality in which one has a UV free electric description when $N_{f}<\frac{3}{2}(N_{c}-2)$.  We map this description to a magnetic description which is weakly coupled in the IR.  The field content of the electric picture is a complex chiral superfield
\begin{center}
\renewcommand{\arraystretch}{1.5}
\begin{tabular}[c]{|c c c c|} \hline 
Field & $SO (N_c) $ & $SU (N_f) $ & $U (1)_R$  \\ \hline 
$Q$ &   {\tiny\yng(1)}  &   {\tiny\yng(1)}  &  $\frac{N_{f}-N_{c}+2}{N_{f}}$  \\ \hline
\end{tabular}
\end{center}
There is also a discrete symmetry associated with  $Q$:
\begin{equation}
Q\rightarrow e^{\frac{2\pi i }{2N_{f}}} Q  \quad N_{c}\ne3  ,\quad Q\rightarrow e^{\frac{2\pi i }{4N_{f}}} Q \quad N_{c}=3. 
\end{equation}
For massless quarks we have $W=0$. At the non-trivial IR fixed point the duality is exact.  It is insightful to note that at the scale invariant fixed point we do not have a well defined particle interpretation, for instance of the gauge bosons of the two dual gauge groups, and the duality between different gauge groups is exact.  Introducing an electric quark mass term to the superpotential
\begin{equation}
W=m \text{Tr}[Q.Q]=m\text{Tr}M
\end{equation}
the duality becomes effective.  
$W=m \text{Tr}[Q.Q]$ introduces a scale and the particle states are well defined: the two theories have a 
different number of gauge bosons and the duality between the two gauge groups is an effective one \cite{Seiberg:1994pq}.


\subsection{Macroscopic theory}
In this section we review the process of supersymmetry breaking for the $SO(N)$ macroscopic (magnetic) theory with $N=N_{f}-N_{c}+4$, which is 
IR free when $N_{f}>3(N-2)$. This is the effective description of an electric theory (microscopic), at energies below the strong coupling scale 
of the macroscopic theory $\Lambda_{m}$.  We may treat the $SO(N)$ gauge symmetry as a global symmetry to extract the vacuum 
symmetries and the field representations and then later gauge this symmetry. The 
electric quark meson $M$ can be related to the magnetic elementary ``meson'' $\Phi$ by $\Phi=M/\Lambda_{e}$ where $\Lambda_{e}$ 
 is the dynamical scale of the electric description.  Further we relate the electric quark masses $m$ to $\mu$ by $\mu^{2}_{ij}=-m_{ij}\Lambda$. 
  $\Lambda$ relates the characteristic scale of the electric description to the magnetic description via $h=\Lambda_{e}/\Lambda$.  

We start with a macroscopic global $SO(N)$, with $N=N_{f}-N_{c}+4$ \cite{intriligator2006dsb}. The field content is 
\begin{center}
\begin{tabular}[c]{|ccccc|} \hline
Field & $SO(N) $&$SU(N_{f})_{f}$& $U(1)' $& $U(1)_{R}$ \\
\hline
$\Phi$ &  $1$& {\tiny\yng(2)} & -2& $2$\\
$\varphi$ & \text{\tiny\yng(1)}  &$\overline{\text{\tiny\yng(1)}}$ &1 &$0$
\\ \hline
\end{tabular}
\end{center}
These are complex chiral superfields. The canonical K\"ahler potential is
\begin{equation}
K=\mbox{Tr}[ \varphi^{\dagger}\varphi]+\mbox{Tr}[\Phi^{\dagger}\Phi].
\end{equation}
The superpotential is 
\begin{equation}
W_{\text{ISS}}=h \mbox{Tr} [\varphi^{T}\Phi \varphi] - h\mu^2 \mbox{Tr} \, \Phi.
\label{1}
\end{equation}
The initial global symmetries are valid when $\mu=0$. Switching on $\mu$ breaks $SU(N_{f})\times U(1)'$ to $SO(N_{f})$:
\begin{center}
\begin{tabular}{|cccc|} \hline
Field & $SO(N) $&$SO(N_{f})_{f}$& $U(1)_{R}$ 
\\
\hline
$\Phi$ &  $1$& {\tiny\yng(2)} &  $2$ \\
$\varphi$ & \text{\tiny\yng(1)}  &\text{\tiny\yng(1)} &$0$
\\
\hline
\end{tabular}
\end{center}
Supersymmetry is broken by rank condition when $N_{f}> N$ and the potential is minimised by 
\begin{equation}
\Phi = \left(
\begin{array}{cc}
0 & 0  \\
0 & X_{0} 
\end{array} \right)\quad  \varphi = \left(
\begin{array}{c}
\chi  \\
0 
\end{array} \right)\quad
\chi = \mu\left(\!
\begin{array}{cc}
\cosh\theta&i\sinh\theta \\
-i\sinh\theta & \cosh\theta
\end{array} \!\right)\otimes  \mathbb{I}_{N/2}
\label{vacs}
\end{equation}
where $X_{0}$ is a $(N_{f}-N)\times (N_{f}-N)$ symmetric matrix and $\chi^{T} \chi =\mu \mathbb{I}_{N}$. We find

\begin{center}
\begin{tabular}{|c|c|c|c|}
\hline
Field & $SO(N)$ & $SO(N)_{f} $ & $SO(N_{f}-N)_{f} $ \\
\hline
$\Phi = \left(
\begin{array}{cc}
Y_{\text{{\tiny $N$x$N$}}} & Z^{T}_{\text{{\tiny $N_{f}$-$N$x$N$}}}  \\
Z_{\text{{\tiny $N$x$N_{f}$-$N$}}} & X_{\text{{\tiny $N_{f}$-$N$x$N_{f}$-$N$}}}
\end{array} \right)_{\text{{\tiny $N_{f}$x$N_{f}$}}} $ &
$ \left(
\begin{array}{cc}
1
\end{array} \right) $& 

$\left(
\begin{array}{cc}
\text{\tiny\yng(2)} & \text{\tiny\yng(1)} \\
\text{\tiny\yng(1)}& 1
\end{array} \right) $ &

$\left(
\begin{array}{cc}
1 & \text{\tiny\yng(1)}  \\
\text{\tiny\yng(1)} &\text{\tiny\yng(2)} + 1
\end{array} \right)$\\

$\varphi=\left ( \begin{array}{c}
\chi_{\text{{\tiny $N$x$N$}}} \\
\rho_{\text{{\tiny $N_{f}$-$N$x$N$}}}
\end{array} \right)_{\text{{\tiny $N_{f}$x$N$}}}$ 

& $\left (\begin{array}{c}
\text{\tiny\yng(1)} \end{array} \right) $ 

&$\left (\begin{array}{c}
\text{\tiny\yng(1)}\\ 1 \end{array} \right)$ 

& $\left(\begin{array}{c}
1 \\
\text{\tiny\yng(1)}
\end{array} \right)$\\
\hline
\end{tabular}
\end{center}
These global flavour symmetries are the symmetries into which we weakly gauge and identify with the standard model GUT ``parent''. The singlet $\text{Tr}X=(1,1,1)$ in the above (where we use notation where $(A,B,C)$ refers to irreps. of $SO(N)\times SO(N)_{f} \times SO(N_{f}-N)_{f} $) is the chiral superfield whose fermionic component is the massless Goldstino arising from the spontaneous breaking of supersymmetry. The pseudo flat directions has degenerate vacuum energy density
\begin{equation}
	V_{min}=(N_{f}-N)|h^2 \mu^4|.
\end{equation}
Perturbative quantum effects create a local minimum at $X=0$, $\chi=\mu \mathbb{I}_{N}$ (up to global symmetries in the parameterisation of
 $\chi$).  The superpotential, in terms of the component fields, is 
\begin{equation}
W= h\text{Tr}[\chi^{T}Y\chi+\rho^{T}X\rho + \chi^{T}Z\rho+\rho^{T}Z^{T}\chi]-h\mu^{2}\text{Tr}  Y- h\mu^{2}\text{Tr}  X.
\label{comp}
\end{equation}
As the global flavour symmetry is broken we can choose to diagonalise $\text{Tr}[\mu^{2}\Phi]$, and assign different values to parameters within each global flavour symmetry group,
\begin{equation}
\mu^2_{AB} = \left(
\begin{array}{cc}
\mu^2 \mathbb{I}_{N} & 0  \\
0 & \hat\mu^2 \mathbb{I}_{N_F-N} 
\end{array} \right)_{A B}
\label{t}
\end{equation}
with $\mu>\hat{\mu}$.  This choice of values is an explicit breaking of the global symmetry and would remove the Goldstone bosons of the vacuum. 
The unbroken symmetry of the vacuum is $SO(N)_{D}\times SO(N_{f}-N)\times U(1)_{R}$:

\begin{center}
\begin{tabular}{|c|c|c|}
\hline
Field & $SO(N)_{D} $& $SO(N_{f}-N)_{f} $\\
\hline
$\Phi = \left(
\begin{array}{cc}
Y_{\text{{\tiny $N$x$N$}}} & Z^{T}_{\text{{\tiny $N_{f}$-$N$x$N$}}}  \\
Z_{\text{{\tiny $N$x$N_{f}$-$N$}}} & X_{\text{{\tiny $N_{f}$-$N$x$N_{f}$-$N$}}}
\end{array} \right)_{\text{{\tiny $N_{f}$x$N_{f}$}}} $ &

$\left(
\begin{array}{cc}
\text{\tiny\yng(2)} & \text{\tiny\yng(1)} \\
\text{\tiny\yng(1)}& 1
\end{array} \right) $ &

$\left(
\begin{array}{cc}
1 & \text{\tiny\yng(1)}  \\
\text{\tiny\yng(1)} & \text{\tiny\yng(2)}+1
\end{array} \right)$\\

$\varphi=\left ( \begin{array}{c}
\chi_{\text{\tiny $N$x$N$}} \\
\rho_{\text{\tiny $N_{f}$-$N$x$N$}}
\end{array} \right)_{\text{\tiny{$N_{f}$x$N$}}}$ 

&$\left (\begin{array}{c}
\text{\tiny\yng(1)}\times \text{\tiny\yng(1)}\\ \text{\tiny\yng(1)}\end{array} \right)$ 

& $\left(\begin{array}{c}
1 \\
 \text{\tiny\yng(1)}
\end{array} \right)$\\
\hline
\end{tabular}
\end{center}
$SO(N)_{D}$ is a colour-flavour locking phase \cite{Rajagopal:2000wf}.  The vacuum has five sectors of fields under equivalent representations. Each sector satisfies $\text{Str} M^{2}=0$. 
The fermionic particles are in Weyl multiplets and the chiral superfields are complex.  These give two real boson mass eigenstates 
for each complex degree of freedom.  These sectors are:

\paragraph{\textbf{Tr} $\mbf{X}$ } The trace of $X$ is the massless Goldstino of the spontaneously broken global supersymmetry and it is accompanied by two real bosons.

\paragraph{$\mbf X$} This field is the classically massless ``pseudo-modulus'' which is one loop lifted by the Coleman Weinberg potential.  For pure ISS (no deformations) its vev is lifted to the origin.  There are $(N_{f}-N)(N_{f}-N+1)-2$ real bosons and half as many Weyl fermions.

\paragraph{$\mbf{ (\rho, Z )}$ } These fields are the largest contribution towards messengers of the soft terms.  There are $2 N (N_{f}-N)$ Weyl fermions coming from ($\psi_{\rho},\psi_{Z}$).  For small vevs of $X$ their mass is approximately $h\mu$.  We will explore their mass in more detail for specific models later.  
The broken global symmetries arising from $SO(N_{f})_{f}\rightarrow SO(N_{f}-N)_{f}\times SO(N)_{f}$ are Goldstone bosons in $Re(\rho)$. Explicit breaking by high dimension operators results in pseudo Nambu-Goldstone Goldstone bosons \cite{Banks:2009cx}.
If one makes an explicit choice of different quark masses (\refe{t}) these Goldstone bosons are avoided. The remaining complex scalars 
generate masses of order $m^{2}\sim h^{2}\mu^{2}$  and there is multiplet splitting between them due to the vev of $X$ and the explicit R symmetry breaking deformations, as expected from a general Heisenberg Wigner mode.

\paragraph{$\mbf{ (Y ,\chi_{S}) } $}  There are $N(N+1)-2$ chiral superfields whose fermions and scalar components have mass of order $h\mu$. As 
   is explored in the next section, increasing the vevs of fields will increase the masses.  Introducing explicit R-symmetry breaking terms  that generate a vev for $Y$ will cause multiplet splittings of the scalar components of these fields.  As $F_{Y}=0$ and $F_{\chi}=0$, these fields play no role as messengers.
   
\paragraph{$\mbf{ \chi_{A}}$}  The antisymmetric part of $\chi$ parameterises the Goldstone bosons and  pseudo-moduli of $SO(N)\times SO(N)_{f}\rightarrow SO(N)_{D}$. If all the electric quark masses are the same $\mu\sim m_{Q}\Lambda$, there are $\frac{N}{2}(N-1)$ complex chiral superfields of which half are Goldstone and the other half pseudo-moduli.  Using the vacuum \refe{vacs} we can
label $\theta_{+}=\theta+\theta^*$ as the pseudo-modulus and $\theta_{-}=\theta-\theta^{*}$ the Goldstone boson. The Coleman Weinberg potential will generate a mass for $\theta_{+}$.  The gauge fields will all acquire mass from the super-Higgs mechanism when $SO(N)_{c}$ is completely gauged, in particular the Goldstone superfields become Higgs superfields which are then eaten by the vector superfields. 
\paragraph{}
As the super-Higgs mechanism is a supersymmetry preserving mechanism, the supertrace holds and there is no effect on the stability of the  vacuum from gauging $SO(N)_{c}$.  The masses of the gauge superfields and the pseudo-moduli superfield are obtained from the D-term potential, giving $m= gh\mu$.  

When we come to consider embedding the standard model in $SO(10)$ flavours  groups, and SO(10) is broken, these particles will form 
irreducible representations of the standard model gauge group, which we discuss in a later section.


\subsection{Choice of embeddings}
There are two choices of standard model embeddings via $SO(10)$.  In the first embedding we put the standard model into the global symmetry group $SO(N_{f}-N)_{f}$. 
 In the second case we embed into the global symmetry $SO(N)_{f}$. The vacuum of the theory has the symmetry  $SO(N)_{D}\times SO(N_{f}-N)_{f}$. 
 Correspondingly, beta function contributions are from the matter's representation under the $SO(N)_{f}$ gauge group.
 
 The gauged $SO(N)_{c}$ is infrared free for $N_{f}>3(N-2)$. The electric description is asymptotically free for $N_{f}<\frac{3}{2}(N_{c}-2)$. 
 If we identify $SO(N_{f}-N)_{f}=SO(10)$ and weakly gauge, then we find $N_{f}-N=10$ and $N=N_{f}-N_{c}+4$ leads to $N_{c}=14$.  These conditions 
 are met when  $18\geq N_{f}\geq12$ with the corresponding N, with $8\geq N\geq 2$.

Alternatively, we may identify $SO(N)_{f}=SO(10)$, then $N_{f}>24$ and this gives $N_{c}>18$.  Of course we may also embed $SO(10)$ into a subset 
of a larger flavour symmetry group.  Identifying the full flavour group is merely the minimal choice when exploring these models.  For a recent example of embedding into a subset, in particular embedding $SU(5)$ into a weakly gauged $SU(6)$ flavour symmetry, see \cite{Franco:2009wf}.


\section{Contribution to gaugino masses}
ISS models generally have a signature of heavy squarks and light gaugino masses \cite{Abel:2007jx,Abel:2007nr,Abel:2008gv}. These mass types 
are soft terms generated from the messengers of the hidden sector.  In this section we will focus on the gaugino masses, as being lighter, they 
pose the initial phenomenological concern.

The vector superfield of $SO(10)$ is in the antisymmetric $\mathbf{45}$ (adjoint), which is traditionally broken at an energy scale, $M_{GUT}$,
 far above the SUSY breaking scale.  The extra gauge degrees of freedom are ``super-Higgsed'', whereby the scalar and fermionic components of a Higgs
  chiral superfield is eaten by a massless vector supermultiplet.  Thus the broken gauge bosons and the gauginos both have masses at the $M_{GUT}$ scale.
   We are thus left with the issue of the gaugino masses of the standard model. Regardless of this Super-Higgs mechanism, all the gauginos 
   of the standard model parent gauge group will get equal contributions to their masses from the messengers. 

ISS models have multiple messengers.  The fundamental messengers  ($\rho,Z$) are the major contributor to gaugino masses, $M_{\gl}$.
These have been 
the focus of much of the literature of gauge
 mediation \cite{Martin:1997ns,Dubovsky:1999xc,Martin:1996zb,Giudice:1998bp,Giudice:1997ni,Dimopoulos:1996ig,Raby:2007yv}. 
 Their contribution to gaugino and sfermion masses depends on the vev of $X$ which is non-zero only if R-symmetry breaking 
 deformations are added to the ISS model, and depends on $F_{X}$. Explicit examples will be discussed in later sections \ref{Gaugino and} and \ref{Gaugino and2}. 
 The other contributions are  either from $X$ or from ($Y,\chi_{S}$).  A concise method for calculating this contribution is outlined 
 in ``General Gauge Mediation'' \cite{Meade:2008wd}. The gaugino mass contribution is calculated from the two point function of the fermionic 
 component of the gauge current superfield. The interested reader may follow appendix A, where we review these techniques in the light of $SO(10)$.

First we calculate the gaugino mass contribution from the X pseudo-modulus. When identifying $SO(N_{f}-N)_{f}=SO(10)$, $X$ is in the symmetric representation of $SO(10)$. 
Schematically the scalar mass squared matrix is:
\begin{equation}
\left(
X^{\dagger} X
 \right) M_{X}^{2}\left(
\begin{array}{c}
X \\
X^{\dagger}
\end{array} \right) \quad \quad  M_{X}^{2} \propto \frac{h^{4}}{64\pi^{2}}\left(
\begin{array}{cc}
\mu^{2}& \braket{X}^{2}  \\
\braket{X}^{2} & \mu^{2}
\end{array} \right)\otimes \mathbb{I}_{N_{f}-N}\otimes\mathbb{I}_{N_{f}-N}
\end{equation}
where we have stripped all indices. The diagonal and off diagonal terms are both found from computing the one loop Coleman Weinberg potential. The diagonal components arise 
from the pure ISS superpotential. The off diagonal components are proportional to $|W_{\rho\rho}|^{4}$, where $W_{\rho\rho}$ is the double
 derivative of the superpotential with respect to $\rho$.  This type of term is classically zero in the ISS model but may appear at one loop, 
 if there are deformations to the ISS model. This matrix will have two mass eigenstates ($m_{X\pm}$). There is one fermionic mass eigenstate
  ($M_{\psi_{X}}$).  Computing the two point function and using Ward identities (see Appendix A) we find, 
\begin{alignat}{1}
M^{X}_{\gl_{r}}&=2M_{\psi_{X}}{g_r}^{2}R[X]T(\mb{r})(D(x;m_{X+})-D(x;m_{X-}))D(x;M_{\psi_{X}})\nonumber \\
&=2M_{\psi_{X}}{g_r}^{2}R[X]T(\mb{r})\int \frac{d^{4}k}{(2\pi)^{4}}(\frac{1}{k^{2}+m^{2}_{X+}}-
\frac{1}{k^{2}+m^{2}_{X-}})\frac{1}{k^{2}+M^{2}_{\psi_{X}}}\nonumber \\
&\propto (\frac{2\alpha_r}{4\pi})R[X]T(\mb{r})\frac{h^{2}\braket{X}^{2}}{8\pi \mu}  \quad \quad \quad  \text{for}\quad  \braket{X}^{2}<\mu^{2}
\label{sols}
\end{alignat}
where $T(\mb{r})$ is related to the Dynkin index of the representation $\mb{r}$, in this case symmetric of $SO(10)$, $T(sym)=12$. $R[X]$ is the rank of the field $X$. The subscript $r$ on the coupling $g_{r}$ denotes the gauge group associated to each coupling, such as $g_{3}$ of $SU(3)$.  The fermionic mass $M_{\psi_{X}}$ may found by taking the $\text{STr}M^{2}=0=M^{2}_{+}+M^{2}_{-}-2M^{2}_{\psi}$ and obtain the 
fermionic mass from the known scalar masses. The result for the pure ISS model is 
\begin{equation}
M_{\psi_{X}}=\frac{h^{4}\mu^{2}}{64{\pi}^{2}}.
\end{equation}
We now look at the ($Y,\chi_{s}$) sector. When identifying $SO(N)_{D}=SO(10)$, these fields are charged under the standard model GUT parent and we should expect them to
 behave as messengers also.  We see that as the superpotential is of the form 
\begin{equation}
W_{ISS}\supset h\chi^{T}Y\chi
\end{equation}
these fields do not behave like the fundamental messengers which had a coupling of the form $W_{ISS}\supset \rho^{T}X\rho$ where $X$ was
 just a background superfield. In this case we would apply the methods \cite{Dimopoulos:1996ig} for multi-messengers where the fermion messenger 
 mass matrix parameterised by the superfields ($Y,\chi_{s}$) is 
\begin{equation}
W= 
\left(
\chi^{T} Y^{T}
 \right) M \left(
\begin{array}{c}
\chi \\
Y
\end{array} \right) \quad \quad  M  =\left(
\begin{array}{cc}
h\braket{Y}& h\mu  \\
h\mu & 0 
\end{array} \right)\otimes \mathbb{I}_{N}\otimes\mathbb{I}_{N}.
\end{equation}
$\chi$ achieves a vev when $F_{Y}=0$ as this is the breaking by rank condition.  For $Y$ to obtain a vev at the minimum requires 
setting $F_{\chi}=0$ in general. 
 The result is that these fields, although possibly charged under the standard model GUT group, cannot generate gaugino or sfermion contributions.  
One may  speculate that a complicated deformation of ISS with magnetic quarks may give a vev to $Y$ and achieve $F_{\chi}\ne0$.   
 In that case we would apply the methods \cite{Dimopoulos:1996ig} for multi-messengers and there would be no supression of gaugino masses at first order in $F$, despite a zero in the fermion mass matrix.÷\footnote{The interested reader may note that the zero 
  in the fermion mass matrix for ($Y,\chi_{s}$) can be filled by a multitrace deformation of the magnetic meson $\text{Tr}\Phi^{2}$.}
For pure ISS the values of the masses and vevs for each field may be found in  \cite{intriligator2006dsb} and are compiled for
 comparison later, in Table 1.
\begin{table}
\begin{center}
\begin{tabular}{|c|c|c|c|c|}
\hline
Field &Tree level vev &One loop vev &$M^{2} $&$M_{\gl }$ \\
\hline
$X$&Moduli&$0$&$\frac{h^{4}\mu^{2}(log 4-1)N(N_{f}-N)}{2\pi^{2}}$&$0$\\
$Y$&0&0&$4 h^{2}\mu^{2}$&0\\
$\chi$&$\mu$&$\mu$&$4h^{2}\mu^{2}$&0\\
$\theta_{+}=(\theta+\theta^{*})$&Moduli&$0$&$\frac{h^{4}\mu^{4}(log 4-1)N(N_{f}-N)}{8\pi^{2}}$&0\\
$\rho,Z$&0&0&$4 h^{2}\mu^{2}$&0\\
\hline
\end{tabular}
\end{center}
\caption{Vevs and masses of scalars fields in pure ISS model. }
\end{table}

 In summary we see that the breaking of an explicit global R-symmetry, in the hidden sector, leads to the explicit soft 
 terms which break the global supersymmetry in the visible sector. These R-symmetry breaking terms control the vev of $Y$ and $X$. We 
 will classify the deformations of ISS model by their influence on the vevs of these fields.
 

\section{Spontaneous versus explicit R-symmetry breaking}

R-symmetry of the superpotential prevents gaugino mass terms from the messengers of the magnetic 
description of the ISS model.  We must then include R-symmetry breaking terms.  The key motivation of ISS models is that they 
satisfy Seiberg dual descriptions at the IR fixed point. So the deformations should be from irrelevant operators, that do not add 
new degrees of freedom (new fields) into the superpotential. The general approach is to add irrelevant operators to the electric
 description which will be parametrically suppressed.

In $SU(N)$ models one has a choice between spontaneous and explicit R-symmetry breaking when adding deformations to the model\footnote{See also an example of spontaneous R-symmetry breaking at two loops  \cite{Amariti:2008uz}.}. For spontaneous R-symmetry breaking one 
requires that some of the fields in the ISS model are $R\ne0$ but that the superpotential still has $R[W]=2$. 

The first term in the ISS superpotential is
\begin{equation}
W_{ISS}\supset h \text{Tr}[\tilde{\varphi} \Phi \varphi]
\end{equation}
where $\tilde{\varphi}$ may signify the antifundamental of $\varphi$ in the $SU(N)$ case, or simply transpose in the $SO(N)$ and $Sp(N)$ cases. 
If we also consider the $\text{Tr}\Phi$ linear term in $W_{\text{ISS}}$ we obtain two constraints
\begin{equation}
R[\tilde{\varphi}]+R[\Phi]+R[\varphi]=2 \quad R[\Phi]=2  \rightarrow R[\tilde{\varphi}]=-R[\varphi].
\end{equation}
In $SO(N)$ and $Sp(N)$ models the $\tilde{\varphi}$ signifies transpose such that the constraints can only ever be satisfied by $R[\varphi]=0$. 
 So we see the only explicit R-symmetry may be used for $SO(N)$ and $Sp(N)$ models.\footnote{However, see Section 4 of \cite{Intriligator:2007py} for a spontaneous breaking of $U(1)_{R}$ for  $SO(N)$ involving $D$ terms and breaking of the $SO(N)$ symmetry for an O'Raifeartaigh model.}

In $SO(N)$ we can use the invariant two index Kronecker ($\delta_{\alpha \beta}$) and Levi-Civita ($\epsilon_{\alpha_{1}...\alpha_{N}}$) symbol 
to build terms that explicitly break R-symmetry using the dual magnetic quarks, explicitly the $\chi$ components of the magnetic quarks $\varphi$. For $SO(N)=SO(2)$ we may have
\begin{equation}
	W= h k\delta_{s t}\delta_{\alpha \beta}\varphi^{s}_{\alpha}\varphi^{t}_{\beta}+  h m\epsilon_{\alpha \beta}\epsilon_{s t}\varphi^{s}_{\alpha}\varphi^{t}_{\beta} 
	\label{2}
\end{equation}
where the $r,s,t$ indices are of $SO(2)_{mag}$ and the $\alpha, \beta$ are from $SO(N)_{f}$. 
This is the deformation used in \cite{Abel:2007jx,Abel:2007nr,Abel:2008gv} for $SU(5)$ models. Baryon deformations of this type will give a 
vev to the $Y$ field.  This will effect the scalar masses of ($Y,\chi$) but in general will have \emph{no effect on gaugino masses}, when embedding 
into either $SO(N)_{D}$ or into $SO(N_{f}-N)_{f}$. 


\subsection{Tree level potential for SO(2) model}

Let us first analyse the dual quark deformations from \refe{2}. Setting $N=2$ and $N_f-N = 10$,  we take \refe{1} and \refe{2} and compute 
the tree level potentials for $SO(2)_{mag}\times SO(2)_{f}\times SO(N_{f}-N)_{f}$
\begin{equation}
	W=  h Tr \varphi^{T}\Phi \varphi - h\text{Tr}[\mu^2 \Phi] + h k\delta_{s t}\delta_{\alpha \beta}\varphi^{T}_{s \alpha}\varphi_{\beta t}+ 
	 h m\epsilon_{\alpha \beta}\epsilon_{s t} \varphi^{T}_{s \alpha}\varphi_{\beta t}. 
\end{equation}
Using the $SO(2)_{f}\times SO(10)_{f}$ symmetry we may diagonalise the matrix $\mu^2_{AB}$ as
\begin{equation}
\mu^2_{AB} = \left(
\begin{array}{cc}
\mu^2 \mathbb{I}_{2} & 0  \\
0 & \hat\mu^2 \mathbb{I}_{10} 
\end{array} \right)_{A B}.
\end{equation}
The resulting tree level potential is
\begin{alignat}{1}
V_{F}&= \sum_{\alpha \beta}|h \chi^{T}_{s \alpha} \chi_{\beta s} - h \mu^2 \delta_{\alpha \beta}|^{2} +  \sum_{a \beta} |2 h \rho^{T}_{s a}\chi_{\beta s}|^{2}\nonumber \\  
& + \sum_{a b} | h \rho^{T}_{s a} \rho_{b s} - h \hat{\mu}^2 \delta_{a b}|^{2}+ \sum_{s a}| 2 h {Z}_{a \beta} \chi_{\beta s} + 2 h X_{a b} \rho_{b s}|^{2} \\
& + \sum_{s \alpha} |2 h Y_{\alpha \beta} \chi_{\beta t} + 2 h Z^T_{\alpha b}\rho_{b s}+ 2h k \delta_{\alpha \beta} \chi_{\beta s} + 2h m \epsilon_{\alpha \beta} \chi_{\beta s}|^{2}. \nonumber
\end{alignat}
The indices are $A=$($\alpha,a$) and $B=$($\beta,b$) running over all $N_{f}$ with $\alpha,\beta$ running over the first $N$ and $a,b$ 
running over the $N_{f}-N $. 
 We apply the rank condition method of ISS and set $\rho=0$.  The potential becomes
\begin{alignat}{1}
V_{F}&= \sum_{\alpha \beta} |h \chi^{T}_{s \alpha} \chi_{\beta s} - h \mu^2 \delta_{\alpha \beta}|^{2} + \sum_{a b} | h \hat{\mu}^2 \delta_{a b}|^{2}
 + \sum_{s a}| 2 h {Z}_{a \beta} \chi_{\beta s}|^{2} \nonumber \\
& + \sum_{s \alpha} | 2h Y_{\alpha \beta} \chi_{\beta t} + 2h k \delta_{\alpha \beta} \chi_{\beta s} + 2h m \epsilon_{\alpha \beta} \chi_{\beta s}|^{2} .
\end{alignat}
The potential is minimised when
\begin{equation}
\chi^T_{s \alpha} \chi_{\beta s}=\mu^2 \mathbb{I}_{\alpha \beta}
\end{equation}
where $\alpha,\beta$ run over $N$.  For $SO(N)$ models $\chi^T_{s \alpha}$ is just the transpose of $\chi_{\beta s}$, they are not 
independent fields. To minimise, we set $Z=0$. Notice also that $V_{F}$ is independent of $X_{ab}$ and these are the pseudo-moduli.  $SO(10)$ is unbroken.  Preliminarily we choose the local minimum to be
\begin{equation}
\braket{\chi} = \Sigma \left(
\begin{array}{cc}
1& 0  \\
0 & 1
\end{array} \right)
+
\theta \left(
\begin{array}{cc}
0& -1  \\
1 &0
\end{array} \right)
\end{equation}
with the constraint 
\begin{equation}
 \Sigma^{2}+\theta^{2}=\mu^2.
\end{equation}
The $\chi$ fields have excitations that are constrained to live on a circle.  The symmetric and antisymmetric piece both preserve the 
$SO(2)_{D}$ symmetry.  We have F-term  constraints on $Y$ to minimise.
 We set $\braket{Y_{\alpha\beta}}=\eta \mathbb{I}_{\alpha\beta}$. Putting this back into the scalar potential we find
\begin{equation}
V_{F}\!=\!2N h^{2}|\eta\Sigma \!+\! m\Sigma \!+\! k\Sigma|^2\!+\! N h^{2}| \!-\! \eta\theta\!-\! m\theta \!-\! k\theta|^2\!+\!N h^{2}|\eta\theta \!+\! m\theta \!+\! k\theta|^2 \!+\! (N_{f}-N)h^2\hat{\mu}^4
\end{equation}
where we have in mind that $N=2$ and $N_{f}-N =10$.  Using the constraint reduces $V_{F}$ to 
\begin{equation}
	V_{F}=2N|h\eta \mu +h k\mu+ h m\mu|^2+(N_{f}-N)h^2 \hat{\mu}^4.
\end{equation}
Minimising in $\eta$ we find
\begin{eqnarray}\label{pot2}
\eta &= &-(k+m) \nonumber \\
V(\hat{\mu})&=&(N_{f}-N)h^2\hat{\mu}^4.
\end{eqnarray}
So the minimum of the scalar potential is independent of the particular choices of $\Sigma$ and $\theta$, with these fields constrained to live on 
a circle of radius $\mu$ in field space.  Choosing a particular value of $\Sigma$ and $\theta$ will break this continuous symmetry. It is clear now 
that the Kronecker contracted and Levi-Civita contracted terms act equivalently to the scalar potential and we may drop one of them without loss of 
generality. For all values of the potential, it is positive definite and non zero in terms of the $Y$($\eta$) field.

It is useful to compare this with the $SU(N)$  ISS models \cite{Abel:2007jx,Abel:2007nr,Abel:2008gv}. In \emph{those} models there is a runaway direction,
 associated with the parameterisation of the vevs of $Y$, $\chi$ and  antifundamental $\bar{\chi}$ fields, which is one loop lifted.  In \emph{those} models 
 the deformation will be significant to the ($\rho,Z$) messenger contributions to gaugino masses and plays an important role when embedding into both 
 flavour groups.


\section{The KOO deformation}
In this section we keep the delta contracted deformation of the previous section, which is valid for any  $SO(N)$ and not just $SO(2)$.  We add to this a new
 deformation. As has previously been pointed out in \cite{Kitano:2006xg}, that to obtain gaugino mass contributions from the fundamental
  messengers ($\rho$, $Z$) at first order in $F$, one must add a deformation that adds a mass term to the diagonal of the messenger mass matrix.  It
   is an explicit R-symmetry breaking term.  We will refer to this as the KOO deformation:
\begin{equation}
W_{KOO}=h^{2}m_{z}\text{Tr} (Z^{T}Z).
\label{koo}
\end{equation}
The full potential we would like to analyse is
\begin{equation}
W= h \mbox{Tr} [\varphi^{T}\Phi \varphi] - h \mbox{Tr} [\mu^2 \Phi] + h k\delta_{s t}\delta_{\alpha \beta}\varphi^{T}_{s \alpha} \varphi_{\beta t} + h^{2}m_{z}\text{Tr} [Z^{T}Z].
\end{equation}
The scalar potential is 
\begin{alignat}{1}
V_{F}&= \sum_{\alpha \beta}|h \chi^{T}_{s \alpha} \chi_{\beta s} - h \mu^2 \delta_{\alpha \beta}|^{2} + \sum_{a \beta} |2 h \rho^{T}_{s a}\chi_{\beta s} 
+ 2 h^2 m_z Z^T_{\beta a} |^{2} \nonumber\\  
&+ \sum_{a b} | h \rho^{T}_{s a} \rho_{b s} - h \hat{\mu}^2 \delta_{a b}|^{2}
  + \sum_{s a}| 2 h {Z}_{a \beta} \chi_{\beta s} + 2h X_{a b} \rho_{b s}|^{2}\\
& + \sum_{s \alpha} |2 h Y_{\alpha \beta} \chi_{\beta t} + 2 h Z^T_{\alpha b}\rho_{b s}+ 2h k \delta_{\alpha \beta} \chi_{\beta s}|^{2}.\nonumber
\end{alignat} 
We can follow the usual steps of minimising the potential.  We find an ISS type minimum, $V(\hat{\mu})=(N_{f}-N)|h^{2}\hat{\mu}^{4}|$ 
with 
\begin{equation}
\braket{\rho}=0 \quad \braket{Z}=0 \quad \braket{Y}=-k.
\end{equation} $X$ is a modulus of the classical potential. The other SUSY broken minimum are found to be 
\begin{alignat}{1}
&\chi^{s}_{\alpha}\chi^{s}_{\beta}=\mu^2 \mathbb{I}_{N} \nonumber\\
&\rho^{s}_{a} \rho^{s}_{b}=\frac{h^2 m^{2}_{z}}{\mu^2}ZZ=\text{diag}(\hat{\mu}^{2}...\hat{\mu}^{2},0...0)_{N_{f}-N}\nonumber\\
&X_{ab}=-\frac{\mu^2}{m_{z}}\text{diag}(1...1,\hat{x}...\hat{x})_{N_{f}-N}\\
&Y_{\alpha\beta}=-(\frac{\hat{\mu}^{2}}{m_{z}}+k)\mathbb{I}_{N}\nonumber\\
&V_{low}=(N_{f}-N-n)|h\hat{\mu}^{2}|^{2}.\nonumber
\label{v1}
\end{alignat}
The label $n$ runs from $1$ to $N$.  The first term has $\hat{\mu}^2$ for the first $N$ corresponding to the rank condition.  The remaining $N_{f}-2N$ of $N_{f}-N$ 
are zero and the $\hat{\mu}^2$ of that F term generate $V_{low}$. The $\hat{x}$ signify classical moduli. These extra SUSY broken minimum only arise 
with the KOO deformation because this deformation gives an extra degree of freedom to fix the $Z$ minimum in the scalar potential. The deformation 
fills the zero of the scalar mass matrix giving gaugino masses at order first order in $F_{X}$. It is also an explicit R-symmetry breaking term, giving a nonzero vev 
to the pseudo-modulus $X$, which, as we discussed above, is crucial.  
Certainly the ISS vacuum is viable, however some speculation over the viability of this secondary and lower minimum is necessary. 
We suggest that it is viable as an $SO(N_{f}-N-n)$ or $SU(N_{f}-N-n)$ embedding of the standard model and give a review of this for $SU(N_{f}-N-n)$ in appendix B.


\subsection{Messenger masses with KOO deformation}
As we have stated already, the deformations using the magnetic quarks (e.g. $hk\delta\delta \varphi\varphi$) does not effect the mass matrix 
of ($\rho,Z$) and does not effect the vev of $X$. The KOO deformation of \refe{koo} does effect both of these quantities. In this section we 
examine the messenger sector of ($\rho,Z$).  For a general set of fields the fermionic mass matrix is given by \cite{Barbieri:1982nz}
\begin{equation}
m^{2}_{1/2} =  \left(
\begin{array}{cc}
W^{ab}W_{bc}+2D^{\alpha a}D^{\alpha}_{c}&-\sqrt{2}W^{ab}D^{\beta}_{b}  \\
-\sqrt{2}D^{\alpha b}W_{bc} & 2D^{\alpha c}D^{\beta}_{c}
\end{array} \right)
\end{equation}
where $W_a = \partial W / \partial \Phi_a$, $W^a = \partial W^\dagger / \partial \Phi^\dagger_a$.
The scalar mass matrix is 
\begin{equation}
m^{2}_{0} =  \left(
\begin{array}{cc}
W^{ab}W_{bc}+D^{\alpha a}D^{\alpha}_{c}+D^{\alpha a}_{c}D^{\alpha}&W^{abc}W_{b}+D^{\alpha a}D^{\alpha c}  \\
W_{abc}W^{b}+D^{\alpha}_{a}D^{\alpha}_{c} &W_{ab}W^{bc}+D^{\alpha}_{a}D^{\alpha c}+D^{\alpha c}_{a}D^{\alpha}
\end{array} \right).
\end{equation}
The D terms are taken to be switched off.  To compute the matrices for ($\rho,Z$), we choose to parameterise the fermion mass matrix by 
\begin{equation}
	\psi=(\rho_{a s },Z_{a \beta}).
\end{equation}
The scalar squared mass matrix is parameterised by $(\rho_{a s },Z_{a \beta}, \rho^*_{a s },Z^*_{a \beta})$.
We make the choice of vevs to be 
\begin{alignat}{3}
\braket{X_{ab}}&=X_0 \mathbb{I}_{ab}   &\quad    \braket{Y_{\alpha\beta}}&=\eta \mathbb{I}_{\alpha\beta} &\quad 
\braket{\chi_{\beta s}}&= \mu \mathbb{I}_{\beta s}.
\end{alignat}
We have switched off the $\theta$ dependence of the $\chi$ vev to achieve analytic results. Hence, for both embeddings the fermion mass matrix is given by
\begin{equation}
m_{f}=\mathbb{I}_{N_{f}-N} \otimes \mathbb{I}_{N}\otimes 2 h \left ( \begin{array}{cc}
X_0&  \mu  \\
\mu & h m_{z}
\end{array} \right).
\end{equation}
The two fermionic eigenvalues are
\begin{equation}
M_{\pm}=h(hm_{z}+X_0\pm\sqrt{4\mu^2+(-hm_{z}+X_0)^2}).
\end{equation}
The scalar mass matrix  is then computed to be
\begin{equation}
m_s^2 = \left( 
\begin{array}{cc}
 W^{ab} W_{bc} &  W^{abc}W_b \\
 W_{abc} W^b & W_{ab} W^{bc}
 \label{massmatrix}
\end{array}
\right) 
\end{equation}
where
\begin{equation}
W^{ab} W_{bc}  = \mathbb{I}_{N_{f}-N} \otimes\mathbb{I}_{N} \otimes 4 h^2 \left( \begin{array}{cc}
X_0 {X_0}^* + \mu^2 & \mu ({X_0}^* + h m_z)  \\
 \mu (X_0+h m_z) & h^2 m_{z}^2 + \mu^2
\end{array} \right)
\end{equation}
and 
\begin{equation}
W^{abc} W_{b} = W_{abc} W^{b} = \mathbb{I}_{N_{f}-N} \otimes  \mathbb{I}_{N} \otimes 2 h^2 \left ( \begin{array}{cc}
-\hat{\mu}^2  & 0  \\
0 & 0 
\end{array} \right).
\end{equation}
The four independent scalar mass squared eigenvalues are
\begin{eqnarray} \label{scalarmasses}
m^{2}_{1,\pm}\! &=& \! h^2 \big[ -\hat{\mu}^{2}\!+\!4\mu^2\!+\!2 h^2m^{2}_{z}\!+\!2|X_0|^2\! \nonumber \\
&&  \hspace{0.5cm} \pm \sqrt{16 \mu^2 (h m_Z+X_0)(h m_Z + {X_0}^*) +(2 h^2 m_Z^2 -(- \hat{\mu}^2 +2 X_0 {X_0}^*))^2} \; \big] \nonumber  \\
m^{2}_{2,\pm} \!&=&\! h^2 \big[ \hat{\mu}^{2}\!+\!4\mu^2\!+\!2 h^2m^{2}_{z}\!+\!2|X_0|^2\! \nonumber \\
&&  \hspace{0.5cm} \pm \sqrt{16 \mu^2 (h m_Z+X_0)(h m_Z + {X_0}^*) +(2 h^2 m_Z^2 -(\hat{\mu}^2 + 2 X_0 {X_0}^*))^2} \; \big].
\end{eqnarray}
We calculate the corresponding Coleman-Weinberg potential
\begin{equation}
V_{CW} = \frac{1}{64\pi^{2}}\text{STr}\mathbf{M}^4\text{Log}\frac{\mathbf{M}^{2}}{\Lambda^{2}}=
\frac{1}{64\pi^{2}}(\text{Tr}\mathbf{m}^{4}_{B}\text{Log}\frac{\mathbf{m}_{B}^{2}}{\Lambda^{2}}-
\text{Tr}\mathbf{m}^{4}_{F}\text{Log}\frac{\mathbf{m}_{F}^{2}}{\Lambda^{2}})
\end{equation}
for the messenger correction at one loop and eventually, find
\begin{equation}
X_{0}=\braket{X} = \frac{1}{2} h m_z, \; \; \; \; M^{2}_X = \frac{h^{4}\hat{\mu}^{2}}{12\mu^{2}\pi^{2}}\left(
\begin{array}{cc}
\hat{\mu}^{2}&  -\frac{9}{40}{X_0}^{2}  \\
-\frac{9}{40}{X_0}^{2} &\hat{\mu}^{2}
\end{array} \right)
\end{equation}
where we have expanded to first order in $h, m_z$ and in $\hat{\mu} / \mu$ up to first non-vanishing order. We have supressed factors of $N(N_{f}-N)$
in the expression for $M^{2}_X $ coming from tracing over degenerate mass eigenavlues.


\subsection{Gaugino and squark masses}
\label{Gaugino and}

Generalisations of the wavefunction renormalisation technique, in the regime that the $F$-term of 
the pseudo-modulus is smaller than the messenger scale, give analytic expressions for the gaugino and sfermion masses
 \cite{Cheung:2007es}. For the gauginos  
\begin{equation}
m_{\gl_r}=\frac{\alpha_r}{4\pi}\Lambda_{G}=\frac{\alpha_r}{4\pi}F_{X}\sum_{i}\frac{\partial_{X}M_{i}}{M_{i}}
\label{gauginos111}
\end{equation}
where $M_{i}$ are the eigenvalues of the fermion mass matrix. The gaugino masses from ($\rho,Z$) are found to be 
\begin{equation}
m^{(\rho,Z)}_{\gl_r}=\frac{\alpha_r h^{2}\hat{\mu}^{2}m_{z}}{2\pi( \mu^2-h m_{z} X_{0})}
\label{gaugino}
\end{equation}
where to simplify expressions we have defined $X_0 = \braket{X} $.
The sfermions from ($\rho,Z$) are given by
\begin{alignat}{1}
m^{2}_{\tilde{f}}&=2\sum_{r}C^{r}_{\tilde{f}}(\frac{\alpha_{r}}{4\pi})^{2}\Lambda^{2}_{S}\nonumber \\
\Lambda^{2}_{S}&=\frac{1}{2}|F_{X}|^{2}\sum_{i}|\frac{\partial_{X}M_{i}}{M_{i}}|^{2}
\label{sfermions111}
\end{alignat}
where $\alpha_{r}$ is the gauge coupling at the messenger scale and $C^{r}_{\tilde{f}}$ denotes the quadratic Casimir of
 the irrep $\tilde{f}$ in the gauge group factor labelled by $r$. We find 
\begin{equation}
\Lambda^{2}_{S,(\rho,Z)}=\frac{(h^2\hat{\mu}^4)[h^4 m_{z}^4 + 2 {\mu}^4 - 2 h^3 m_{z}^3 X_{0} - 2 h m_{z} {\mu}^2 X_{0} + 
  h^2 m_{z}^2 (4 \mu^2 + X_{0}^2)]}{[(\mu^2 - h m_{z} X_{0})^2 (4 \mu^2 + (-h m_{z} + X_{0})^2)]}
\end{equation}
The KOO deformation $h^2m_{z} \mbox{Tr} [Z^{T}Z]$ is a mass term for some of the messengers of the theory.  As is highlighted in
 \cite{Cheung:2007es}, the introduction of messenger masses changes the ratio 
\begin{equation}
 \frac{\Lambda^{2}_{G}}{\Lambda^{2}_{S}}=N\rightarrow N_{eff}(h,\mu,m_{z},X_{0})
\end{equation}
where $N_{eff}$ is the effective messenger number, which can continuously vary from $0$ to $N$ inclusive. 
\begin{equation}
\frac{\Lambda^{2}_{G}}{\Lambda^{2}_{S}}=\frac{4 h^2 m_{z}^2 [4 \mu^2 + (-h m_{z} + X_{0})^2]}{[
 h^4 m_{z}^4 + 2 \mu^4 - 2 h^3 m_{z}^3 X_{0} - 2 h m_{z} \mu^2 X_{0} + 
  h^2 m_{z}^2 (4 \mu^2 + X_{0}^2)]}.
\end{equation}
In our case it ranges from 0 to 4 as expected (see for instance \cite{Abel:2009ze}).  The messenger field $X$ makes the contribution to gaugino masses 
\begin{equation}
m^{X}_{\gl_r}=\frac{2\alpha_r}{4\pi}T(\mb{r})R[X]\sqrt{3}[\frac{3h^{2}\hat{\mu}^{2} X_{0}^{2}}{80\pi\mu^{3}}]
\end{equation}
and for the sfermions 
\begin{alignat}{1}
m^{2}_{\tilde{f},X}&=2\sum_{r}C^{r}_{\tilde{f}}(\frac{\alpha_{r}}{4\pi})^{2}T(\mb{r})R[X]\Lambda^{2}_{S,X} \nonumber \\
\Lambda^{2}_{S,X}&=\frac{27 h^4 \hat{\mu}^4  X_{0}^4}{6400 \mu^6 \pi^2}.
\end{alignat}


\subsection{Constraints on parameters}
It is useful to constrain the parameter space for comparison between differing models.  A more accurate constraint on the parameters would involve a more detailed phenomenological survey of the model.  We review closely, the constraints used in \cite{Zur:2008zg}. 

This model has a dimensionless coupling $h$, and has five parameters  $(\mu,\hat{\mu},m_{z},k,\Lambda_{m})$.
\begin{itemize}

\item $\frac{h }{4\pi}$ is used for a perturbative expansion; we require h at $O(1)$.

\item Cosmological bounds on the gravitino mass give
\begin{alignat}{1}
m_{3/2}&=\frac{F}{\sqrt{3}M_{pl}}<16 eV\\
h\mu^2&=\frac{F}{\sqrt{N_{f}-N}}<\frac{\sqrt{3}16 M_{pl}eV}{\sqrt{N{f}-N}}<(146TeV)^2.
\end{alignat}
$M_{pl}$ is the reduced Planck mass.  This is of the same order as \cite{Zur:2008zg} namely $(174TeV^2)$.

\item We can determine $\hat{\mu}$ as the ratio of $\frac{\mu}{\hat{\mu}}$ which controls the longevity of the metastable vacuum from 
\begin{equation}
S\sim(\frac{\mu}{\hat{\mu}})^{4}(\frac{\mu}{m_{z}})^{4}.
\end{equation}

\item The scalar masses are completely equivalent to those found in appendix A of \cite{Zur:2008zg}.  In particular there is a ``no Tachyon'' constraint 
\begin{equation}
|\mu^2\pm hm_{z}X|^2>\hat{\mu}^2(\mu^2+h^{2}m^{2}_{z}).
\end{equation}
This will give a constraint on $m_{z}$ dependent on the values of h, $\mu$ and $\hat{\mu}$.

\item We can determine the cutoff scale $\Lambda_{m}$ from the longevity of the metastable vacuum to the non-perturbative SUSY vacuum. Normally we expect that
\begin{equation}
|\epsilon|= |\frac{\mu}{\Lambda_{m}}| \ll 1 
\end{equation}
is sufficient suppression of tunneling to the non-perturbative vacuum. In cases where there is no large hierarchy between 
$\Lambda_m$ and $\mu$,(e.g in order to avoid low energy Landau poles in SM gauge coupling constants) then it is suffices to take $\hat{\mu}$ 
sufficiently smaller than $\mu$ to avoid too short a lifetime of the metastable vacuum.  Numerical examples are given later.
\item  The term $k$ plays no role in either gaugino masses or vacuum stability. It plays no significant role to visible sector phenomenology.
\end{itemize}



\section{The non-perturbative potential}

In this section we explore the non-perturbative potential of the macroscopic theory.  This is perturbative in the electric theory. 
 In general one finds that when gauging of the $SO(N)_{c}$ symmetry group and taking the model to be IR free when $N_{f}>3(N-2)$, 
 one finds a non-perturbative potential 
\begin{equation}
W_{dyn}=(N-2)(h^{N_{f}}{\Lambda}_{m}^{3(N-2)-N_{f}}\text{det}\Phi)^{1/(N-2)}.
\end{equation}
The supersymmetry preserving vacuum is found at 
\begin{equation}
\braket {\varphi}=0 \quad \braket{\Phi}=\frac{\mu}{h} (\frac{\Lambda_{m}}{\mu})^{\frac{N_{f}-N+2}{3N-N_{f}-4}}\mathbf{1}_{N_{f}}.
\end{equation}
When $\epsilon=\frac{\mu}{\Lambda_{m}}<<1$ the metastable vacuum will be exponentially long lived.  This formula being valid when the electric 
quark masses are all equal.  In general the vev of $X$ and $Y$ is fixed by solving $\pd{1}{W}{X}= \pd{1}{W}{Y}=0, \braket {\varphi}=0 $ 
where the superpotential is the full classical plus dynamically generated part.  We find
\begin{alignat}{1}
\braket{X}&=h^{\frac{N_{f}-2N +4}{N-N_{f} -2}}\hat{\mu}^{\frac{4}{N-N_{f}-2}}\mu^{-\frac{2N}{N-N_{f}-2}}\Lambda_{m}^{-\frac{(N_f-3(N-2))}{N-N_{f}-2}}\nonumber \\
&\nonumber \\
\braket{Y}&=h^{\frac{N_f-2N+4}{N-N_f-2}}\hat{\mu}^{\frac{2(N-N_f )}{N-N_f-2}}
\mu^{\frac{2(-2N +N_{f}+2)}{N-N_{f}-2}} \Lambda_{m}^{-\frac{(N_f-3(N-2))}{N-N_{f}-2}}.
\end{alignat}

The above SUSY minimum applies to the case $N \neq 2$.
For the specific embedding where $N=2$,$N_{f}=12$ has $N_{f}=N_{c}-2$ so the macroscopic (IR) theory is in the coulomb phase \cite{intriligator2006dsb}. 
The IR ISS superpotential should be multiplied by an arbitrary function $f(t)$ where 
\begin{equation}
 t=Det (\Phi)/\Lambda^{24} 
\end{equation}
subject to the boundary condition $f(0)=1$. To leading order in $\Phi$ the SUSY broken vacuum is independent of this function.  The magnetic $SO(2)_{c}$ is Higgsed and the unbroken electric $SO(2)_{c}$ is 
confined. Thus we have a metastable SUSY broken vacuum in a confining phase.  For a more detailed exploration of these and other cases one can look at the 
original papers \cite{intriligator2006dsb,intriligator1995dmd}.


\subsection{Lifetime of the metastable vacuum}
The ISS vacuum can decay into either the secondary SUSY broken minimum or into the SUSY restored non-perturbative vacuum.  In general one can apply 
the techniques outlined in \cite{Duncan:1992ai}. Here we review some analytic estimates applicable to this model when tunneling into the SUSY
restored minimum. At the ISS minimum
\begin{equation}
V_{ISS}=(N_{f}-N)|h\hat{\mu}|^{2}.
\end{equation}
The value of the pseudo-modulus is found from the Coleman Weinberg potential to be
\begin{equation}
X_{0}=\braket{X}=\frac{hm_{z}}{2}.
\end{equation}
We can estimate the value of the local maximum of the potential by expanding the superpotential around the vevs $Z=\rho=0$. There is a local
 maximum when $\tilde{\chi}\chi=0$:
\begin{equation}\label{approx2}
V_{Peak}=N|h\mu^2|^{2}+(N_{f}-N)|h\hat{\mu}|^{2}\approx N_{f}|h\mu^{2}|^{2}
\end{equation}
We can also estimate the value of the pseudo-modulus at this vev by use of the Tachyon constraint
\begin{equation}
X_{0}=\braket{X}=\frac{\mu^2-\hat{\mu}\sqrt{\mu^2+h^{2}m^{2}_{z}}}{h m_{Z}}
\end{equation}
We use the triangle approximation\cite{Duncan:1992ai} when $\mu\sim\hat{\mu}$:
\begin{equation}
S\approx \frac{(\Delta X)^{4}}{V_{peak}}\approx \frac{(\braket{h\Phi})^{4}}{V_{peak}}=\frac{h^{2}}{ N_{f}}(\frac{\Lambda}{\mu})^{4(\frac{N_{f}-N+2}{3N-N_{f}-4})}
\end{equation}
When $\mu\ne\hat{\mu}$ we find that:
\begin{equation}
S\approx  \frac{1}{N_f} h^{-\frac{ (6N-2N_f-12)}{ N- N_f -2}}{\epsilon '}^{\frac{16}{N-N_f-2}} {\epsilon }^{-\frac{(12N-4N_f-8)}{N-N_f-2}} 
\end{equation}

where we have defined  $\epsilon' = \hat{\mu}/{\Lambda_m}$ and $\epsilon = {\mu}/{\Lambda_m}$. In deriving the above expression we have assumed that 
$\mu $ and $\hat{\mu}$ though unequal  are approximately the same order, so that (\ref{approx2}) is still a good approximation.

One can compare this with S=400 for the age of the universe. Typically the vevs of the fields are small compared to the distance 
from the origin to the SUSY restored vacuum. Also the $V_{peak}$ value is in general independent of the deformations used. For
 these reasons, this result is consistent with the deformations used in this paper. To derive the actual value of the lifetime we will need to 
 input the various allowed values of $\Lambda_m$, $\mu$ and $\hat{\mu}$ consistent with various other conditions which we will consider later. 

When using the KOO deformation one can also tunnel from the ISS to the second SUSY broken minimum. For this particular model, with the baryon 
deformation switched off, the calculation is completely equivalent to that carried out in appendix B of \cite{Zur:2008zg}, in which the calculation 
was numerical. In \cite{Kitano:2006xg} there is an analytic estimate of the bounce action, using the triangle approximation: 
\begin{equation}
S \sim (\frac{\mu}{\hat{\mu}})^{4}(\frac{\mu}{m_{z}})^{4}.
\end{equation}
For a clear discussion of metastability and the thermal favourability of the ISS SUSY broken minimum we highlight \cite{Jaeckel:2008jp}.


\section{Particles of the supersymmetry breaking sector}

The pseudo-modulus $X$ is much lighter than the $m^{2}\sim h^{2}\mu^{2}$ messenger scale.  Some of the 
 irreducible representations of $X$ will be charged when decomposed under the standard model.  These irreps make for a useful signature of ISS type models in collider experiments. For $SO(N_{f}-N)_{f}=SO(10)$, $X_{ab}$ is a symmetric traceless two
  index tensor where the trace is the Goldstino.  Under a $SU(5)$ and standard model $SU(3)_{c}\times SU(2)_{L}\times U(1)_{Y}$ embedding 
  we would find for $X$\cite{Slansky:1981yr}:
\begin{eqnarray}
\mathbf{54}&= \mathbf{15}_{S}\oplus\mathbf{\bar{15}}_{S}\oplus\mathbf{24}_{A} \nonumber \\
\mathbf{15}&= (6,1,-\frac{2}{3})\oplus (3,2,\frac{1}{6}) \oplus (1,3,1)\nonumber \\
\mathbf{\bar{15}}&= (\bar{6},1,\frac{2}{3})\oplus (\bar{3},2,-\frac{1}{6}) \oplus (1,3,-1)\nonumber \\
\mathbf{24}&= (8,1,0)\oplus (1,3,0) \oplus (3,2,-\frac{5}{6})\oplus (\bar{3},2,\frac{5}{6})\oplus(1,1,0)
\end{eqnarray}
where we have chosen to normalise $U(1)_{Y}=\frac{1}{6} U_{1}$ as in  \cite{Slansky:1981yr}.
Under the left right symmetric model $SU(4)_{C}\times SU(2)_{L}\times SU(2)_{R}$  One arrives at
\begin{equation}
\mathbf{54}= (1,1,1)\oplus (6,2,2) \oplus (20',1,1)\oplus (1,3,3).
\end{equation}
Under $SU(3)_{c}\times SU(2)_{L}\times SU(2)_{R}\times U(1)_{Y}$  the above irreps become
\begin{eqnarray}
(1,1,1)&= &(1,1,1,0) \nonumber \\
(6,2,2)&= &(3,2,2,\frac{2}{3})\oplus(\bar{3},2,2,-\frac{2}{3})\nonumber \\
(20',1,1)&=&(\bar{6},1,1,-\frac{4}{3})\oplus(6,1,1,\frac{4}{3})\oplus(8,1,1,0) \nonumber  \\
(1,3,3)&=&(1,3,3,0)
\end{eqnarray}
In general the direct gauge mediation models  with different embeddings and different deformations can somewhat adjust the parameters and 
phenomenological predictions.  Still better though is a clear feature that distinguishes between different hidden sector models.  The relatively
 light mass of the pseudo-modulus is a good window into the hidden sector. For example, we can compare this to the $SU(5)$ case explored in \cite{Zur:2008zg}.  
 In those models the pseudo-modulus is an adjoint $\mathbf{24}$ of $SU(5)$.  The $SO(10)$ case has extra matter from the $\mathbf{15}$  and 
  $\bar{\mathbf{15}}$ reps that can distinguish between these two hidden sector gauge groups.  This can be general applied for all
   $SU(N_{f}-N)_{f}$, $SO(N_{f}-N)_{f}$ and $SP(N_{f}-N)_{f}$ models: A different weakly gauged flavour group will predict different 
   particle content due to the light pseudo-modulus.


\subsection{Beta function contributions to the visible sector}
We can associate the parent of the standard model with either of the two flavour groups.  If we choose to embed in the $SO(N_{f}-N)_{f}=SO(10)$, 
 then the fields ($X,\rho,Z$) contribute to the beta functions of both standard model $SU(3)_{c}$ and $SU(2)_{L}$:
\begin{equation}
\delta b_{2,3}(\mu_{R}<2h\mu)= -12 \quad \delta b_{2,3}(2h\mu<\mu_{R}<\Lambda)= -12-2N \quad  \delta b_{2,3}(\mu_{R}>\Lambda)=-N_{c}
\end{equation}
The delta signifies that these contributions should be added to the visible sector beta functions of the MSSM or SUSY GUT models.
For the embedding of $SO(N)_{f}$,  the contributions to the beta functions are 
\begin{equation}
 \delta b_{2,3}(\mu_{R}<2h\mu)=0 \quad \delta b_{2,3}(2h\mu<\mu_{R}<\Lambda)= -12-N_{f} \quad \delta b_{2,3}(\mu_{R}>\Lambda)=-N_{c}.
\end{equation}
For the SM and MSSM one has the standard results,
\begin{equation}
b^{SM}_{a}=(-41/10,19/6,7) \quad b^{MSSM}_{a}=(-33,-1,3).
\end{equation}
These additional contributions to  the beta functions agree with other direct gauge mediation models explored:  embedding into $SO(N_{f}-N)_{f}$ results in a Landau pole before 
the unification scale. This is clear from the large contribution of $-12$ to $b_3$ from the $X$ fields for $\mu_{R}<2h\mu $ 
which results in a Landau pole in the strong coupling $\alpha_3$ at $\mu_R  \sim $\, 10TeV.

On the other hand embedding the standard model gauge group into $SO(N)_{f}$ can avoid a Landau pole by making the messenger masses significantly heavy.
Specifically we find that for $N=10, N_f =26$ (recall for this embedding we must have $N_f > 24$) values of $2h\mu  \sim 10^{14}$ GeV and 
$\Lambda \sim 10^{15}$ can  just avoid a pole below 
$M_{gut} \sim 2.2\times 10^{16}$ GeV. This implies a value of  the parameter $\epsilon \sim \frac{1}{20h} $. Thus to establish a long enough lifetime for 
the metastable vacuum is apparently more stringent in this case than in $SU(N)$ based ISS models.  


\subsection{Gauging both flavour groups}
Simultaneous gauging of both flavour groups is a novel feature of $SO$ models. In the case of orthogonal groups a natural semi-direct product group that contains the standard model is
 $SO(6) \times SO(4) $ with the $SO(6)$ containing as a  maximal subgroup $SU(3)_c \times U(1)$ and 
$SO(4)\sim SU(2)_L \times SU(2)_R $. Starting from our macroscopic flavour group $SO(N)_D \times SO(N_f-N)_f $ 
there might appear 2 ways to gauge $SO(6)\times SO(4) $ namely  either choosing $N=4, N_f=10$ or $N=6, N_f=10 $

The constraints coming from infrared freedom of the $SO(N)_c$ and asymptotic freedom of the electric theory requires
$N_f > 3(N-2) $ and $N_f < \frac{3}{2} (N_c-2) $, with $N = N_f-N_c+4 $. 
Of the 2 possible embeddings only the choice $N_f=10, N=4$ (so $N_c=10$ ) satisfies both these constraints.  We notice that in the electric picture these two global symmetries arise from the same $SO(N_{f})=SO(10)$ and make for an interesting  possible unification in the electric theory where a unification in the magnetic picture would appear doubtful. The gaugino mass and sfermion mass contributions from the  $(\rho,Z)$ messengers would be the same for both flavour groups (up to R.G. flow) as has been calculated in this paper except that the contributions from the $X$ messenger would be solely to $SO(6)$. The contributions to the beta functions to $SO(6)$ would be
\begin{equation}
\delta b_{SO(6)}(\mu_{R}<2h\mu)= -12 \quad \delta b_{SO(6)}(2h\mu<\mu_{R}<\Lambda)= -20 
\end{equation}
and for $SO(4)$
\begin{equation}
 \delta b_{SO(4)}(\mu_{R}<2h\mu)=0 \quad \delta b_{SO(4)}(2h\mu<\mu_{R}<\Lambda)= -22.
\end{equation}
The presence of the $-12$ in the beta function of $\alpha_3$ means a Landau pole is present at low energies 
(around 10TeV) so unfortunately this maximal, simultaneous gauging of both flavour groups has problematic phenomenology. 
However in may be that by studying the `reduced rank' version of ISS based on SO groups (discussed later in subsection 8.7) 
different macroscopic flavour groups emerge that might still allow for simultaneous gauging of both flavour groups in such 
a way as to reduce the number of light messenger fields that produced the low energy Landau pole above.


\section{Multitrace deformations}
In this section we examine introducing multitrace deformations of the magnetic quarks $\varphi$ and the meson field $\Phi$ \cite{Giveon:2007ef,Essig:2008kz,Xu:2007az}. The deformations of magnetic quarks mix fields of the two global symmetry groups in which one may embed the standard model GUT.  These operators have been suppressed by the strong coupling scale of the magnetic picture. Whilst we do not supply a UV completion we find them useful as they demonstrate how the vev of the pseudo-modulus may be shifted without changing the $F_{X}$. In contrast, the KOO deformation demonstrated a shifting of $\braket{X}$ and a filling of a zero of the messenger fermion mass matrix. This zero had suppressed gaugino masses at first order in $F_{X}$.  Finally, multitraces of the meson field are explored. These deformations are irrelevant operators of $Q$ from the UV electric description which are then mapped to $\Phi$ of the magnetic description. These deformations will make the $F_{X}$ term $X$ dependent and the KOO deformation naturally appears as part of this meson deformation.  As the scalar potential is no longer independent of $X$, it is no longer a classically flat direction although one loop corrections still contribute to its minimisation.


\subsection{Multitrace deformation of magnetic quarks}
We may build a deformation out of magnetic quarks of the form
\begin{equation}
\delta W= \frac{h\epsilon}{\Lambda_{m}}( \mbox{Tr} [\varphi^{T}\varphi])^{2}.
\end{equation}
We keep the KOO deformation and add
\begin{eqnarray}
\delta W &=& h\eta \left[(\chi_{\alpha s} \chi_{s \beta}^T \chi_{\beta t} \chi_{t \alpha}^T ) + ( \chi_{\alpha s} \rho_{s b}^T \rho_{b t} \chi_{t \alpha}^T ) +
 (\rho_{a s} \chi_{s \beta}^T \chi_{\beta t} \rho_{t a}^T ) + (\rho_{a s} \rho_{s b}^T \rho_{b t} \rho_{t a}^T ) \right] \nonumber \\
&+& h \gamma  \left[ (\chi_{\alpha s} \chi_{s \alpha}^T) (\chi_{\gamma t} \chi_{t \gamma }^T) +  (\chi_{\alpha s} \chi_{s \alpha}^T) (\rho_{c t} \rho_{t c}^T ) +
 (\rho_{a s} \rho_{s a}^T ) (\chi_{\gamma t} \chi_{t \gamma }^T) + (\rho_{a s} \rho_{s a}^T ) (\rho_{c t} \rho_{t c}^T )\right]. \nonumber \\
 &&
\end{eqnarray}
Calculating the derivatives of the deformation gives
\begin{equation}
\frac{\partial \delta W}{\partial \chi_{\beta s}} = h\eta [4 \chi_{\alpha s} \chi_{\beta t} \chi_{t \alpha}^T + 4 \chi_{\beta s} \rho_{s a}^T \rho_{a t}] + 
h\gamma [ 4 \chi_{\beta s }(\chi_{\alpha t }\chi^{T}_{t \alpha }) + 4 \chi_{ \beta s} (\rho_{c t} \rho_{t c}^T) ]
\end{equation}
and similarly
\begin{equation}
\frac{\partial \delta W}{\partial \rho_{b t}} = h\eta \left[4 \rho_{b s} \chi_{s \alpha}^T \chi_{\alpha t} 
+ 4 \rho_{b s} \rho_{s a}^T \rho_{a t} \right] + 
h\gamma \left[ 4 (\chi_{\alpha s} \chi_{s \alpha}^T )\rho_{t b}^T +
 4 \rho_{b t}(\rho_{c s}\rho^{T}_{s c}) \right].
\end{equation}
The rest of the  F-terms are the same as for the KOO-superpotential.

In minimising F terms we take
\begin{equation}
 \braket{\rho}=\braket{Z}=0\quad \chi^{T}_{s\alpha}\chi_{\beta s}=\mu^{2}\mathbb{I}_{\alpha \beta} \quad \braket{Y_{\alpha\beta}}=-2(\eta+\gamma)\mu^{2}\mathbb{I}_{\alpha\beta}.
\end{equation} The scalar potential is 
\begin{equation}
V_{min}=(N_{F}-N)|h\hat{\mu}^{2}|^{2} .
\end{equation}
There is a local maximum with $\chi=0$ and Y undetermined at
\begin{equation}
V= (N_{F}-N)|h\hat{\mu}^{2}|^{2}+N|h\mu^{2}|^{2}
\end{equation}
The fermion and scalar mass matrices are 
\begin{equation}
	m_{f}=\mathbb{I}_{N_{f}-N} \otimes \mathbb{I}_{N}\otimes 2 h \left ( \begin{array}{cc}
X_0+2\mu^{2}\Delta& \mu \\
\mu & h m_{z}
\end{array} \right)
\end{equation}
where we have defined  $\Delta=(\eta+\gamma) $. In particular we find that $F_{\chi}^\dagger =0$ and $\pd{3}{W}{\rho}\propto \rho=0$. Using \refe{massmatrix} we find
\begin{equation}
W^{ab} W_{bc}  = \mathbb{I}_{N_{f}-N} \otimes\mathbb{I}_{N} \otimes  4 h^2 \left( \begin{array}{cc}
|X_0\!\!+\! 2\mu^{2}\Delta|^{2} \!+ \!|\mu|^{2}\!\!&  \!\!\mu({X_0}^{*}\!+\!2\mu^{2}\Delta\!+\!\!hm_{z}) \!\!  \\
\mu(X_0\!+\!2\mu^{2}\Delta\!+\!\!hm_{z})  & |\mu|^{2}\!+\!|hm_{z}|^{2} \!\!  
\end{array} \right)
\end{equation}
and
\begin{equation}
W^{abc} W_{b} = W_{abc} W^{b} = \mathbb{I}_{N_{f}-N} \otimes  \mathbb{I}_{N} \otimes 2 h^2 \left ( \begin{array}{cc}
-\hat{\mu}^2  & 0  \\
0 & 0 
\end{array} \right).
\end{equation}
The independent mass eigenvalues for the scalar fields  are
\begin{alignat}{1}
m^{2}_{1,\pm}&=  h^2 ( - \hat{\mu}^2 + G) \pm \sqrt{(G - \hat{\mu}^2 )^2 +H }\nonumber \\
&\nonumber \\
m^{2}_{2,\pm}&=   h^2 ( \hat{\mu}^2 + G ) \pm \sqrt{(G + \hat{\mu}^2)^2 +H }
\end{alignat}
where the functions $G $ and $H$ are defined as 
\begin{eqnarray}
G &=& 8 \Delta^2 \mu^4+ 2 X_0 {X_0}^{*} + 2 h^2 m_{z}^2+4 \mu^2 (1 + \Delta X_0 + \Delta {X_0}^{*}) \nonumber \\
H &=& -8 (2 \mu^4 (1 - 2 \Delta h m_{z})^2 - h^2 m_{z}^2 (\hat{\mu}^2 - 2 X_0 {X_0}^{*}) \nonumber \\
& & \hspace{2cm}  + \mu^2 (-\hat{\mu}^2 + 2 h m_z (-1 + 2 \Delta h m_{z}) (X_0+ {X_0}^{*}))).
\end{eqnarray}
Expanding the Coleman Weinberg potential  we find the VEV of $X$ to first order in ($\eta+\gamma$) and first order in $m_{z}$, to be
\begin{equation}
X_0= \frac{1}{2}h m_{z}-2 \Delta \mu^{2}.
\end{equation}
The mass matrix for $X$ is given by
\begin{equation}
M^{2}_X = \frac{h^4}{\pi^2} \left(
\begin{array}{cc}
\frac{\hat{\mu}^4}{12 \mu^2 }&  
-\frac{3 \Delta \hat{\mu}^4}{40 \mu^2 }X_0
-\frac{3 \hat{\mu}^4}{160 \mu^4 } {X_0}^2 \\
-\frac{3 \Delta \hat{\mu}^4}{40 \mu^2 }X_0
-\frac{3 \hat{\mu}^4}{160 \mu^4 }  {X_0}^2&\frac{ \hat{\mu}^4}{12 \mu^2}
\end{array} \right).
\end{equation}
These expression are sufficient to determine the contributions to gaugino and sfermion masses.


\subsection{Gaugino and squark masses from multitrace of magnetic quarks}
For this model the fermion mass matrix differs from the KOO case we use \refe{gauginos111} we find the  gaugino masses to be
\begin{equation}
m^{(\rho,Z)}_{\gl_r}=\frac{\alpha_r h^{2}\hat{\mu}^{2}m_{z}}{4\pi( \mu^2(1-2h m_{z}\Delta)-h m_{z} X_{0})}.
\label{gauginos8}
\end{equation}
In exactly the same way we find the sfermion masses from \refe{sfermions111} 
giving
\begin{alignat}{1}
& \Lambda^{2}_{S} =  \\
& = \frac{h^{2}\hat{\mu}^{4} [ (h^{4} m^{4}_{z} \!+\! 2\mu^{4}) \!-\! (2 h^{3} m^{3}_{z} \!+\! 2hm_{z}\mu^{2} ) ( 2 \mu^{2} \Delta \!+\! 
X_{0}) \!+\! h^{2} m^{2}_{z} (4 \Delta^{2} \mu^{4} \!+\! X^{2}_{0} \!+\! 4\mu^{2}(1 \!+\! \Delta X_{0}) ]}{2[ \mu^{2} (-1\!+\! 2 h m_{z}\Delta) \!+\!
h m_{z} X_{0}]^{2} [4\mu^{2} \!+\! (-h m_{z} \!+\! 2 \Delta \mu^{2} \!+\!  X_{0})^{2}]}. \nonumber 
\end{alignat}
Equally,
\begin{equation}
N^{(\rho,Z)}_{eff}=\frac{2 h^2 m^2 (4 u^2\! +\! [-h m\! +\! 2\Delta u^2\! +\!X_{0} ]^2)}{
h^4 m^4\! + \!2 u^4\! - (\!2 h^3 m^3 + 2 h m u^2)[2\Delta u^2\! +\! X_{0}] +\! 
 h^2 m^2 (4 \Delta^2 u^4 \!+\! X_{0}^2\! +\! 4 u^2 [1\! +\! \Delta X_{0}])}.
\end{equation}
The messenger field $X$ makes the contributions to gaugino masses 
\begin{equation}
m^{X}_{\gl_r}=\frac{2\alpha_r}{4\pi} \text{T}(\mb{r}) \frac{3\sqrt{3} h^{2}\hat{\mu}^{2}{\mu^2}X_{0}[4 (\eta +
 \gamma) \mu^2 +X_{0}]}{
  80\mu^3 \pi}
\end{equation}
and for the sfermions 
\begin{alignat}{1}
m^{2}_{\tilde{f},X}&=2\sum_{r}C^{r}_{\tilde{f}}(\frac{\alpha_{r}}{4\pi})^{2} \text{T}(\mb{r}) R[X]\Lambda^{2}_{S,X}, \nonumber \\
\Lambda^{2}_{S,X}&=\frac{27 h^4 \hat{\mu}^4 X_{0}^2 [4 \Delta \mu^2 + X_{0}]^2}{6400 \mu^6 \pi^2}.
\end{alignat}


\subsection{Multitrace deformation of meson field}
\label{mesondeform}
In this section we explore the multitrace deformations suggested in \cite{Giveon:2007ef,Essig:2008kz,Xu:2007az}. First we explore the case that all the electric quark masses are the same.  In the second case we again explicitly split the electric quark masses with the hierarchy $m<<\hat{\mu}<\mu<<\Lambda$.  \footnote{Here, $m$ is playing the role of the $m_{z}$ of the previous models.} The superpotential we explore is 
\begin{alignat}{1}
W&=h \mbox{Tr} [\varphi^{T}\Phi \varphi] -h \mbox{Tr} [ \mu^2 \Phi]+\frac{h^{2}m}{2}  \mbox{Tr} [\Phi^{2}] + \frac{h^{2}m\gamma}{2} \mbox{Tr} [\Phi]^{2}\nonumber\\
&=h[\chi^{T}Y\chi+\rho^{T}X\rho + \chi^{T}Z^{T}\rho+\rho^{T}Z\chi]-h\mu^{2} \mbox{Tr} Y -h\hat{\mu}^{2}\mbox{Tr} X+h^{2}m \mbox{Tr} [Z^{T}Z]\\
&+\frac{h^{2}m}{2}[(\mbox{Tr}[Y^{2}]+\gamma{\mbox{Tr}[Y]}^{2})+(\mbox{Tr}[X^{2}]+\gamma{\mbox{Tr}[X]}^{2})]+\frac{h^{2}m\gamma}{2}\mbox{Tr} X \mbox{Tr} Y.
\nonumber
\end{alignat}
In order to minimise the tree-level potential we take the vevs of the fields to be
\begin{equation}
 \braket{\rho}=\braket{Z}=0, \quad \braket{Y}=0, \quad \braket{\chi}=q_{0},  \quad \braket{X}=X_0
\end{equation} 
and find for the F-terms of the fields $X $ and $Y$
\begin{alignat}{1}
F_{X}^\dagger &=-h\hat{\mu}^{2}+h^2 m (1+\gamma (N_{f}-N))X_0\\
F_{Y}^\dagger &=h q_0^2-h\mu^{2}+h^{2}\frac{m \gamma}{2}(N_{f}-N)X_0.
\end{alignat}
This determines $q_{0}$ to be of the form $h q^{2}_{0}=h\mu^{2}-h^{2}\frac{m \gamma}{2}(N_{f}-N)X_0$. 
From the superpotential we find the fermionic mass matrix to be
\begin{equation}
	m_{f}=\mathbb{I}_{N_{f}-N} \otimes \mathbb{I}_{N}\otimes 2 h \left ( \begin{array}{cc}
X_0&  q_{0}  \\
q_{0}& h m
\end{array} \right).
\end{equation}
with independent eigenvalues
\begin{equation}
M_{\pm}=h(hm_{z}+X_0\pm\sqrt{4\mu^2+(-hm_{z}+X_0)^2}).
\end{equation}
Computing derivatives of the scalar potential one finds the scalar mass matrix to be 
\begin{alignat}{1}
&m^{2}_{s}= 4h^2 \mathbb{I}_{N_{f}-N} \otimes  \mathbb{I}_{N} \otimes  \nonumber \\
& \hspace{1cm} \left( \begin{array}{cccc}
 q_0 q_0^* + X_0 {X_0}^* & h m q_0^* + q_0 {X_0}^* & F_{X}^{\dagger}/ 2 h & 0 \\
h m q_0 + q_0^* X_0 & h^2 m^2 q_0 q_0^* & 0 & 0 \\
 F_{X}/ 2h & 0 & q_0 q_0^* + X_0{ X_0}^* & h m q_0 + q_0^* X_0  \\ 
0 & 0 & h m q_0^* + q_0{X_0}^* & h^2 m^2 + q_0 q_0^*  \;
\end{array} \right).
\end{alignat}
The independent eigenvalues are given by
\begin{alignat}{1}
m^{2}_{1\pm}&= h^2 \bigg\{2 h^2 m^2 + 2 X_0 {X_0}^* + 4 q_{0} q_0^* \pm |\hat{\mu}^{2}- h m X_0 (1 + \gamma (N_{f}-n)) | +  \nonumber \\ 
& + \sqrt{16 |h m q_0 + X_0 q_0^* |^2 + (2 X_0{ X_0}^* - 2 h^2 m^2 \pm |\hat{\mu}^{2}- h m X_0 (1 + \gamma (N_{f}-n)) | )^2}, \; \bigg\} \\
 m^{2}_{2\pm}& =   h^2 \bigg\{ 2 h^2 m^2 + 2 X_0{X_0}^* + 4 q_{0} q_0^* \pm |\hat{\mu}^{2}- h m X_0 (1 + \gamma (N_{f}-n)) | -  \nonumber \\
& + \sqrt{16 |h m q_0 + X_0 q_0^* |^2 + (2 X_0 {X_0}^* - 2 h^2 m^2 \pm |\hat{\mu}^{2}- h m X_0(1 + \gamma (N_{f}-n)) | )^2} \; \bigg\}.
\end{alignat}
The meson deformation results in a scalar potential $V$, dependent on $X$; $X$  is no longer a classically flat direction. In \cite{Giveon:2007ef} The vev of $X$ is found by considering the scalar potential plus $\mbox{Tr}[X X^\dagger]$ terms from the one loop Coleman Weinberg contribution. We initially expanded the Coleman Weinberg potential in $X$ and $X^*$ around $0$ and find linear terms in the expansion:
\begin{equation}
V(X) = V^{\mbox{\tiny tree}}(X) +V^{\mbox{\tiny 1-loop} } (X)  = |F_X|^2 + B X X^* + C|X| + D |X^*|
\end{equation}
where we take $F_Y=0$ for our choices of VEVs. We note that the B term  (one loop mass term) will correspond to the b term of \cite{intriligator2006dsb,Giveon:2007ef,Essig:2008kz}.

\subsection{Equal electric quark masses} 
We first consider the case of equal electric quark masses,  $\hat{\mu}=\mu$, and discuss two scenarios, one where the multitrace deformation is switched off ($\gamma = 0$) and the case with $\gamma\ne 0$. The classical plus one-loop Coleman Weinberg potential to first order in $m$ and second order in $X$ is given by (taking $X$ to be real)
\begin{equation}
V(X) = V^{\mbox{\tiny tree}}(X) +V^{\mbox{\tiny 1-loop} } (X)  = F_X^2 + b \; \frac{h^4 \mu^2}{\pi^2} X^2 + c \; 2 \frac{ h^5 \mu^2 m}{\pi^2} X
\end{equation}
where $F_X, b$ and $c$ are in general $\gamma$ dependent.  There is a factor $N\times (N_{f}-N)$ overall.

\paragraph{The $\mbf{\gamma = 0}$ case:}

Switching off the $\text{Tr}[\Phi]^{2}$ deformation, the coefficients $b$ of the mass term and $c$ of the linear term are given by 
\footnote{ note that in the ${\rm SU}$ based ISS case, $b = Log4 -1$ but in both ${\rm SO }$ and ${\rm SU }$ cases $b>0$}
\begin{eqnarray}
b &=& \log [ (\frac{3^9}{256})^{\frac{1}{4}}] - 1, \\
c &=&  \frac{1}{48} \left(51 + 21 \log [2 h^2 \mu^2] + 48 \log [4 h^2 \mu^2]  - 63 \log [6 h^2 \mu^2]\right).
\end{eqnarray}
We note that we have scaled the coefficients in the full scalar potential by a factor of $1 / \pi^2$ compared to \cite{Giveon:2007ef,Essig:2008kz}. 
This potential gives 
\begin{equation}
X_{0}=\braket{X} = \frac{m \mu^2 - c \; h^2 m \mu^2}{h (b \mu^2 + m^2)} , \; \; \; \; M^{2}_X = \frac{h^{4}}{\pi^{2}} \left(
\begin{array}{cc}
b \mu^2   &   {X_0} f + {X_0}^{2} g  \\
 {X_0} f + {X_0}^{2} g  & b \mu^2 
\end{array} \right).
\end{equation}
We remind the reader that again overall factors of $N(N_{f}-N)$ coming from tracing over degenerate mass eigenvlaues, have been taken out. Also, we have not included tree level mass terms in the diagonal components of $M^{2}_X $ since they are sub leading of order $\mathcal{O}(m^2)$.The off-diagonal components of the mass matrix are given by the two functions
\begin{eqnarray}
f &=& \frac{m h}{24}(172 + 87 \log[2 h^2 \mu^2] + 192 \log[4 h^2 \mu^2] - 279 \log[6 h^2 \mu^2] ) \\
g &=& \frac{1}{24} (86 + 30 \log[2 h^2 \mu^2] + 132 \log[4 h^2 \mu^2] - 162 \log[6 h^2 \mu^2] ).
\end{eqnarray}
The vev of X gets a correction proportional to $c $ due to the inclusion of the linear term into the full potential.

\paragraph{The $\mbf{\gamma \neq 0}$ case:}

We now turn on the $\gamma$ deformation and find that to first order in $m$, the mass term of $X$ does not depend on $\gamma$. The coefficient of the linear term has $\gamma$ dependence,
\begin{eqnarray}
c &=& - \frac{1}{16 }  \bigg( 17 - \gamma (N_f - N) + 7 \log [2 h^2 \mu^2] + 4 (4 - \gamma (N_f - N)) \log [4 h^2 \mu^2]  \nonumber \\
& & \hspace{6.4cm}+ (-21 + 6 \gamma (N_f - N)) \log [6 h^2 \mu^2] \bigg).
\end{eqnarray}
We find 
\begin{eqnarray}
X_{0}=\braket{X} &=& \frac{m \mu^2 (1 + \gamma (N_f - N) - c \; h^2 )}{h (b \mu^2 + m^2 (1 + \gamma (N_F - N))^2)}, \\
M^{2}_X &=& \frac{h^{4}}{\pi^{2}} \left(
\begin{array}{cc}
b \mu^2   &   {X_0} f_\gamma + {X_0}^{2} g_\gamma  \\
 {X_0} f_\gamma + {X_0}^{2} g_\gamma  & b \mu^2 
\end{array} \right),
\end{eqnarray}
where $f_\gamma$ is given by 
\begin{eqnarray}
f_\gamma &=& \frac{h m}{24} \bigg( 172 - 3 (- 29 + \gamma (N_f - N) ) \log [2 h^2 \mu^2] + 12 (16 + \gamma (N_f - N) ) \log [4 h^2 \mu^2] \nonumber \\
& &  \hspace{3cm} - 9 (31 + \gamma (N_f - N) ) \log [6 h^2 \mu^2] \bigg)
\end{eqnarray}
and $g_\gamma = g$. Again, $X_0$ is shifted by a correction proportional to $c $. 

\subsection{Unequal electric quark masses} 
We may now make use of the hierarchy between $\hat{\mu}$ and $\mu$ and expand in $\hat{\mu} / \mu$. The classical plus one loop potential is given by 
\begin{equation}
V (X) =  F_X^2 + \tilde{b} \; \frac{h^4 \mu^2}{\pi^2} X^2 + \tilde{c} \; \frac{2  h^5 \mu^2 m}{\pi^2} X
\end{equation}
and again, $F_X, \tilde{b}$ and $\tilde{c}$ are $\gamma$ dependent.

\paragraph{The $\mbf{\gamma = 0}$ case:}

Switching this particular deformation off, the coefficients of the potential are given by 
\begin{equation}
\tilde{b} = \frac{\hat{\mu}^4}{12  \mu^4} 
 \quad  \tilde{c} = - \frac{ \hat{\mu}^2}{48  \mu^4} (9 \mu^2 + 2 \hat{\mu}^2 + 6 \mu^2 \log [4 h^2 \mu^2] ).
\end{equation}
From this potential we find
\begin{equation}
X_{0} = \frac{ m \hat{\mu}^2 - \tilde{c} \; h^2 m \mu^2}{h (b \mu^2 + m^2)}  \; \; \; \; M^{2}_X = \frac{h^{4}}{\pi^{2}}\left(
\begin{array}{cc}
\tilde{b} \mu^2 &   {X_0} \tilde{f} + {X_0}^{2} \tilde{g}  \\
{X_0} \tilde{f} + {X_0}^{2} \tilde{g}  & \tilde{b} \mu^2
\end{array} \right)
\end{equation}
where 
\begin{equation}
\tilde{f} = \frac{h m \hat{\mu}^4}{40 \mu^4} - \frac{h m \hat{\mu}^2}{12 \mu^2} 
\quad \tilde{g} = - \frac{3 \hat{\mu}^4}{160 \mu^4}.
\end{equation}
Also in the case of unequal electric quark masses the tree-level mass terms of the $X$ field are sub leading and we do not include them in the mass matrix.

\paragraph{The $\mbf{\gamma \neq 0}$ case:}

The coefficient of the linear term in the considered potential is given by
\begin{eqnarray}
\tilde{c} &=& - \frac{ \hat{\mu}^2}{48  \mu^4} \bigg( 9 \mu^2 [1+\gamma(N_f - N)] +\hat{\mu}^2 [2 + \frac{3}{2} \gamma (N_f - N)]\nonumber \\
& & \hspace{4cm} + 6 \mu^2 [1 + \gamma (N_f - N)] \log (4 h^2 \mu^2) \bigg)
\end{eqnarray}
whereas the mass term $\tilde{b}$, stays the same to first order in $m$.
Using the full scalar potential we find for the VEV and the mass matrix 
\begin{eqnarray}
X_{0} &=& \frac{m ( \hat{\mu}^2 [1 + \gamma (N_f - N)]- \tilde{c} \; h^2 \mu^2)}{ h (\tilde{b} \mu^2 + m^2 [1 + \gamma (N_F - N)]^2)}, \\
M^{2}_X &=& \frac{h^{4}}{\pi^2}\left(
\begin{array}{cc}
\tilde{b} \mu^2 &   {X_0} \tilde{f}_\gamma + {X_0}^{2} \tilde{g}_\gamma  \\
{X_0} \tilde{f}_\gamma + {X_0}^{2} \tilde{g}_\gamma  &\tilde{b} \mu^2 
\end{array} \right)
\end{eqnarray}
where $\tilde{g}_\gamma = \tilde{g}$  and $\tilde{f}_\gamma$ is given by
\begin{eqnarray}
\tilde{f}_\gamma &=& \frac{ h m \hat{\mu}^4}{40 \mu^4} - \frac{h m \hat{\mu}^2}{12 \mu^2} [1 + \gamma (N_f - N)] + \frac{h m \hat{\mu}^4}{48 \mu^4} \gamma (N_f - N).
\end{eqnarray}


\subsection{Gaugino and squark masses from multitrace of meson}
\label{Gaugino and2}
The multitrace model has the same fermion mass matrix as the generic model with KOO deformation in the previous sections.  
However the $F_{X}$ terms now differ. Using again \refe{gauginos111} the gaugino masses are
\begin{equation}
m^{(\rho,Z)}_{\gl_r}=\frac{\alpha_r h^{2}m(\hat{\mu}^{2}+hm[-1+\gamma(N-N_{f})]X_{0})}{4\pi( q_{0}^2-h m X_{0})}
\label{gauginos3}
\end{equation}
Similarly using \refe{sfermions111} we find the masses of the sfermions to be
\begin{alignat}{1}
& \Lambda^{2}_{S,(\rho, Z)}= \\
& = \frac{h^2(\hat{\mu}^2\! +\! hm[\gamma(N-N_{f})-1]X_{0})^{2}[h^4 m^4\!  + \! 2 {q_{0}}^4\!  - \! 2 h^3 m^3 X_{0} \! - \! 2 h m {q_{0}}^2 X_{0}\!  + \! 
h^2 m^2 (4 q_{0}^2\! +\!  X_{0}^2)]}{2[(q_{0}^2\!  -\!  h m X_{0})^2 (4 q_{0}^2 + (-h m + X_{0})^2)]}\nonumber
\end{alignat}
and the effective messenger number is then given by
\begin{equation}
N^{(\rho,Z)}_{\text{eff}}=\frac{\Lambda^{2}_{G}}{\Lambda^{2}_{S}}=
\frac{2h^{2}m^{2}[(-hm+X_{0})^{2}+4q_{0}^{2})]}   {[h^{4}m^{4}+2q_{0}^{4}-2h^{3}m^{3}X_{0}-2h mq_{0}^{2}X_{0}+h^{2}m^{2}(X_{0}^{2}+4q_{0}^{2})]}.
\end{equation}
The gaugino masses from the $X$ messenger are 
\begin{equation}
M^{X}_{\gl_r}=T[\mb{r}]R[X](\frac{\alpha_r}{4 h^{2}\pi^{2}})\frac{[X_{0}f+X_{0}^{2}g]}{\sqrt{b\mu^{2}}}
\end{equation}
and the sfermion contribution is calculated using
\begin{equation}
\Lambda^{2}_{S,X}=(\frac{h^{4}}{\pi^{2}})\frac{[X_{0}f+X_{0}^{2}g]^{2}}{b\mu^{2}}.
\end{equation}
For the different cases one may scale ($b,f,g$) to ($\tilde{b},\tilde{f},\tilde{g}$) and switch $\gamma$ on or off as appropriate.


\subsection{Uplifted vacuum}
A recent suggestion  \cite{Giveon:2009yu} to improve the viability of ${\rm SU}$ based ISS models is to reduce the rank of the magnetic quark 
matrix: $\text{Rank}(\varphi^{T}\varphi)=k<N $. These new vacuum are higher in energy than the ISS vacuum of full rank and are 
metastable with respect to decay to the ISS vacuum. In particular, new minimal fundamental messengers (labeled $\omega$) 
are formed that are tachyonic in some range of the parameter space. Their contribution to gaugino masses will somewhat alleviate 
the problem of light gauginos and  heavy sfermion (quantified by $N_{eff}$) usually found in the ISS model  building literature. 
To stabilise $\braket{X}$, deformations are added to the basic `SS with reduced rank' model.  In this section we use meson multitrace operators. 
To stay away from these tachyonic directions it is important that we construct a hierarchy that
\begin{equation}
 m \ll \hat{\mu} < \braket{X} <  \mu \ll \Lambda.
\end{equation}

We shall now consider applying the idea of `uplifting' to our ${\rm SO}$ based ISS models.
To begin, let us first focus back to the field content of the model \refe{1} when we break the symmetry groups with a reduced rank, except now the electric quark mass is 
\begin{equation}
\mu^2_{AB} = \left(
\begin{array}{cc}
\mu^2 \mathbb{I}_{k} & 0  \\
0 & \hat\mu^2 \mathbb{I}_{N_{f}-k} 
\end{array} \right)_{A B}.
\label{t2}
\end{equation}
In this case the rank $k<N$ magnetic quark matrix will break both the $SO(N)_{c}$ and $SO(N_{f})_{f}$. The magnetic quark matrix is $(\varphi^{T}\varphi)_{N_{f}\times N_{f}}$, in which there is a trace over the unwritten indices. Breaking by reduced rank give constraints from the magnetic quark matrix:
\begin{alignat}{1}
(\chi^{T} \chi & + \rho^{T}\rho)_{k\times k}=\mu^{2} \mathbb{I}_{kk} \nonumber \\
(\chi^{T}\sigma &+ \rho^{T}\omega)_{k \times N-k} =  0 \times \mathbb{I}_{k \times N-k}  \nonumber\\
(\sigma^{T}\sigma &+ \omega^{T} \omega)_{N-k\times N-k} =0 \times \mathbb{I}_{N-k\times N-k}.
\end{alignat}
Using the usual VEVs, the middle condition implies that $\braket{\sigma}=0$.  The vacuum energy without deformations is 
\begin{equation}
V=(N_{f}-k)|h^{2} \hat{\mu}^{4}|
\end{equation}
with $X$ again a classically flat direction.   To obtain the full rank breaking and return to an ISS \emph{type} vacuum, one would set 
\footnote{One should not confuse this ISS \emph{type} vacuum with the ISS vacuum that has full rank breaking with the energy V= $(N_{f}-N)|h^{2}{\hat{\mu}}^{4}|$  
and only ($\rho, Z$) fundamental messengers. In the former the gauge symmetry that is completely Higgsed is actually $SO(N-k)\times SO(k)$.}
\begin{equation}
(\omega^{T} \omega)_{N-k\times N-k}=\hat{\mu}^{2} \mathbb{I}_{N-k\times N-k}. 
\end{equation}
which is lower in energy by 
\begin{equation}
\Delta V=(N-k)|h^{2}\hat{\mu}^{4}|.
\end{equation}  
As such, this reduced rank breaking is metastable with regard to the lower minimum. 
We keep the traditional $(\chi^{T}\chi)_{\alpha\beta}=\mu^{2}\mathbb{I}_{\alpha\beta}$ and find  \newline
\begin{center}
\begin{tabular}{|c c c c c|}
\hline
Field & $SO(k)_{c}$  & $ SO(N\!-\!k)_{c}$ & $SO(k)_{f} $& $ SO(N_{f}\!-\!k)_{f}$\\
\hline
$\Phi = \left(
\begin{array}{cc}
Y_{\text{{\tiny k x k}}}  &  {Z}^{T}_{\text{{\tiny k x(Nf-k)}}}  \\
Z_{\text{{\tiny (Nf-k)xk }}} & X_{\text{{\tiny (Nf-k)x(Nf-k)}}}
\end{array} \right)_{\text{{\tiny NfxNf}}} $ 

&
$1$&$1$&

$\left(
\begin{array}{cc}
\text{\tiny\yng(2)}&\text{\tiny\yng(1)}\\
\text{\tiny\yng(1)}& 1
\end{array} \right) $ 
&

$\left(
\begin{array}{ccc}
1& \text{\tiny\yng(1)} \\
\text{\tiny\yng(1)} &\text{\tiny\yng(2)}+1
\end{array} \right)$\\

$\varphi = \left ( \begin{array}{cc}
\chi_{\text{{\tiny k x k}}}&\sigma_{\text{{\tiny k x N-k}}} \\
 \rho_{\text{{\tiny  Nf-k x k}}} &\omega_{\text{{\tiny Nf-k x N-k}}}
\end{array} \right)_{\text{\tiny{NfxN}}}$ 

&$\left (\begin{array}{cc}
\text{\tiny\yng(1)}&1\\
\text{\tiny\yng(1)}&1\\
\end{array} \right)$ 

&$\left (\begin{array}{cc}
1&\text{\tiny\yng(1)} \\
1&\text{\tiny\yng(1)}
\end{array} \right)$ 

& $\left (\begin{array}{cc}
\text{\tiny\yng(1)}&\text{\tiny\yng(1)}\\
1&1
\end{array} \right)$ 
& $\left (\begin{array}{cc}
1&1\\
\text{\tiny\yng(1)}&\text{\tiny\yng(1)}
\end{array} \right)$ \\
\hline
\end{tabular}
\end{center} 
The vacuum of unbroken global symmetries have the field content ($\chi=\mu\mathbb{I}$):
\begin{center}
\begin{tabular}{|cccc|}
\hline
Field  & $ SO(N\!-\!k)_{c}$ & $SO(k)_{D} $& $ SO(N_{f} -k)_{f}$ 
\\
\hline
$\Phi = \left(
\begin{array}{cc}
Y_{\text{{\tiny k x k}}}  & {Z}^{T}_{\text{{\tiny k x(Nf-k)}}}  \\
Z_{\text{{\tiny (Nf-k)xk }}} &X_{\text{{\tiny (Nf-k)x(Nf-k)}}}
\end{array} \right)_{\text{{\tiny NfxNf}}} $ &

$1$ &
$\left(
\begin{array}{cc}
\text{\tiny\yng(2)}&\text{\tiny\yng(1)} \\
\text{\tiny\yng(1)}& 1
\end{array} \right) $ &

$\left(
\begin{array}{ccc}
1& \text{\tiny\yng(1)}  \\
\text{\tiny\yng(1)} &\text{\tiny\yng(2)} +1
\end{array} \right)$\\

$\varphi = \left ( \begin{array}{cc}
\chi_{\text{{\tiny k x k}}} & \sigma_{\text{{\tiny k x N-k}}} \\
 \rho_{\text{{\tiny  Nf-k x k}}} &\omega_{\text{{\tiny Nf-k x N-k}}}
\end{array} \right)_{\text{\tiny{NfxN}}}$ 
&
$
\left (\begin{array}{cc}
1&\text{\tiny\yng(1)} \\
1&\text{\tiny\yng(1)}
\end{array} \right)$ 
& 
$\left (\begin{array}{cc}
\text{\tiny\yng(1)}\times\text{\tiny\yng(1)}&\text{\tiny\yng(1)}\\
\text{\tiny\yng(1)}&1
\end{array} \right)$ 
&
$\left (\begin{array}{cc}
1&1\\
\text{\tiny\yng(1)}&\text{\tiny\yng(1)}
\end{array} \right)$ 
\\
\hline
\end{tabular}
\end{center}
Here, $SO(k)_{c}$ is completely Higgsed and again forms part of a colour flavour locking phase.
The superpotential is 
\begin{alignat}{1}
W=& \; h \mbox{Tr}[\varphi^{T}\Phi \varphi] - h \mbox{Tr}[\mu^{2} \Phi]+\frac{h^{2}m}{2} \mbox{Tr}[\Phi^{2}]+ \frac{h^{2} m \gamma}{2} \mbox{Tr}[\Phi]^{2} \nonumber \\
=& \; h[\chi^{T}Y\chi+\rho^{T}X\rho + \chi^{T}Z^{T}\rho+\rho^{T}Z\chi]-h\mu^{2} \mbox{Tr}Y - h\hat{\mu}^{2} \mbox{Tr} X + h^{2}m \mbox{Tr}[Z^{T}Z] \nonumber \\
&+h[ \sigma^{T}Y\sigma +  \omega^{T} Z \sigma + \sigma^{T} Z^{T} \omega +    \omega^{T}X\omega ]\nonumber\\
&+\frac{h^{2}m}{2}[(\mbox{Tr}[Y^{2}]+\gamma{ \mbox{Tr}[Y]}^{2})+(\mbox{Tr}[X^{2}]+\gamma{\mbox{Tr}[X]}^{2})]+\frac{h^{2}m\gamma}{2}\mbox{Tr} X \mbox{Tr} Y.
\end{alignat}
In the above we have also included the meson multi-trace deformation 
term $[\mbox{Tr}\Phi]^2 $ in addition to $ \mbox{Tr}[\Phi^2]$.
The F terms are given by
\begin{alignat}{1}
F_X^\dagger &= h (\rho^T \rho -  \hat{\mu}^2\mathbb{I} +  \omega^T \omega) + h^2 m X (1 + \gamma (N_f - k)) + \frac{h^2 m }{2} \text{Tr}[Y] \mathbb{I}\gamma k, \nonumber\\
F_Y^\dagger &= h (\chi^T \chi - \mu^2\mathbb{I}  + \sigma^T \sigma ) + h^2 m Y (1 + \gamma k) + \frac{h^2 m }{2}\text{Tr}[ X ]\mathbb{I}\gamma (N_f - k), \nonumber\\
F_\omega^\dagger &= h (2 Z \sigma + 2 X \omega),\nonumber \\
F_\sigma^\dagger &= h (2 Y \sigma + 2 \omega^T Z).
\end{alignat}
Minimising under the condition that $\braket{\omega}=0$, the vevs are
\begin{equation}
\braket{\rho}=\braket{Z}=0, \quad \braket{Y}=0, \quad \braket{\sigma}=0, \quad \braket{\chi}=q_0
\end{equation}
where we have $q_0^2 = - \mu^2 + \frac{h m}{2} \gamma X (N_f - k)$. We now expect two separate sectors to contribute to the one 
loop scalar potential in addition to the tree level potential
\begin{equation}
V (X) = V_{\mbox{\tiny tree}} + V_{\mbox{\tiny 1-loop}}^{(\rho, Z)} + V_{\mbox{\tiny 1-loop}}^{(\omega)}.
\end{equation}

\paragraph{The ($\mbf{\rho, Z}$) sector}
From the uplifted superpotential we find that fermion mass matrix is given by
\begin{equation}
	m_{f}=\mathbb{I}_{N_{f}-k} \otimes \mathbb{I}_{k}\otimes 2 h \left ( \begin{array}{cc}
X_0&  q_0  \\
q_0  & h m
\end{array} \right)
\end{equation}
whereas the bosonic mass matrix squared is
\begin{alignat}{1}
&m^{2}_{s}= \mathbb{I}_{N_{f}-N} \otimes  \mathbb{I}_{N} \otimes 4 h^2 \nonumber \\
& \hspace{1cm} 
\left( \begin{array}{cccc}
 q_0 q_0^* + X_0 {X_0}^* & h m q_0^* + q_0 {X_0}^* & F_{X}^{\dagger}/ 2 h & 0 \\
h m q_0 + q_0^* X_0 & h^2 m^2 q_0 q_0^* & 0 & 0 \\
 F_{X}/ 2h & 0 & q_0 q_0^* + X_0 {X_0}^* & h m q_0 + q_0^* X_0  \\ 
0 & 0 & h m q_0^* + q_0 {X_0}^* & h^2 m^2 + q_0 q_0^*  \;
\end{array}
\right).
\end{alignat}
They are identical to the mass matrices of the deformation by a meson field. 
This is as expected because the uplifted model does not introduce any new mixing between ($\rho$, $Z$) and the
 other fields. Note that now we include a tree level mass term for $X$. The tree level term is roughly of the same 
 order as the one loop contribution. The mass matrix is therefore given by
\begin{eqnarray}
M^{2}_X &=& \frac{h^{4}}{\pi^2}\left(
\begin{array}{cc}
\tilde{b} \mu^2 + m^2 \pi^2 (1+\gamma(N_f-k))^2 &   X_0 \tilde{f}_\gamma + {X_0}^{2} \tilde{g}_\gamma  \\
X_0 \tilde{f}_\gamma + {X_0}^{2} \tilde{g}_\gamma  &\tilde{b} \mu^2  + m^2 \pi^2 (1+\gamma(N_f-k))^2
\end{array} \right)
\end{eqnarray}
for the general case of $\hat{\mu} \neq \mu$ and $\gamma \neq 0$. Again we have used
\begin{eqnarray}
\tilde{b} &=& \frac{\hat{\mu}^4}{12  \mu^4}, \nonumber \\
\tilde{c} &=& -  \frac{ \hat{\mu}^2}{48  \mu^4} \bigg( 9 \mu^2 [1+\gamma(N_f - N)] +\hat{\mu}^2 [2 + \frac{3}{2} \gamma (N_f - N)] \nonumber \\
& &  \hspace{4cm} +6 \mu^2 [1 + \gamma (N_f - N)] \log [4 h^2 \mu^2] \bigg), \nonumber \\
\tilde{f}_\gamma &=& \frac{ h m \hat{\mu}^4}{40 \mu^4} - \frac{h m \hat{\mu}^2}{12 \mu^2} [1 + \gamma (N_f - N)] + \frac{h m \hat{\mu}^4}{48 \mu^4} \gamma (N_f - N), \nonumber \\ 
\tilde{g}_\gamma &=& - \frac{3 \hat{\mu}^4}{160 \mu^4}. \nonumber
\end{eqnarray}
Results for the other cases can be found in the corresponding sections of \ref{mesondeform}. As before, we keep the quadratic and linear terms in $X$.  The result is 
\begin{equation}
V_{\mbox{\tiny 1-loop}}^{(\rho, Z)}(X) = \tilde{b} \; \frac{h^4 \mu^2}{\pi^2} X^2 + \tilde{c} \; \frac{2  h^5 \mu^2 m}{\pi^2} X.
\end{equation}

\paragraph{The $\mbf{\omega}$ sector}
The reduced rank produces new messengers $\omega$. The $\omega$ field does not mix with the other messengers. The fermion mass is 
\begin{equation}
	m_{f}= 2 h X_0
\end{equation}
and parameterising by $(\omega, \omega^*)$  we find
\begin{alignat}{1}
m^{2}_{\omega}&= \mathbb{I}_{N_{f}-k} \otimes  \mathbb{I}_{N-k} \otimes 4 h^2 
\left( \begin{array}{cc}
X_0{ X_0}^* &  \frac{F_X^\dagger}{2 h} \\
\frac{F_X}{2 h} & X_0 {X_0}^* \\
\end{array} \right).
\end{alignat}
The bosonic eigenvalues are 
\begin{equation}
m^{2}_{\omega\pm}= 2 h^2 (2 X_0 {X_0}^* \pm | \hat{\mu}^{2} - hm X_0 (1+\gamma (N_{f}-k)) |).
\end{equation}
In order to express the contribution from the $\omega$ messenger to the full potential we make use 
of the hierarchy $\hat{\mu} <   X_0  < \mu$. We expand the Coleman Weinberg potential for
 the $\omega$ sector and keep only the leading logarithm terms (as in \cite{Giveon:2009yu}) ignoring terms higher than quadratic order in 
 $X$. Expanding the potential and after cancellations, we find
\begin{equation}
V_{\mbox{\tiny 1-loop}}^{(\omega)} = \frac{h^4 \hat{\mu}^4}{8 \pi^2} \log[4 h^2 X^2].
\end{equation}

\paragraph{The scalar potential at one loop}
The full scalar potential is given by
\begin{equation}
V (X) = F_X^2 +  k \tilde{b} \; \frac{h^4 \mu^2}{\pi^2} X^2 + k \tilde{c} \; \frac{2  h^5 \mu^2 m}{\pi^2} X + (N-k) \frac{h^4 \hat{\mu}^4}{8 \pi^2} \log[4 h^2 X^2]
\end{equation}
There are $(N_{f}-k)$ copies of this potential coming from the trace on $X$.  This plays no role on the minimisation as it is an overal factor on both sides. $F_X^2$ is given by
\begin{equation}
F_X^\dagger = - h\hat{\mu}^2 + h^2 m X (1 + \gamma (N_f-k)).
\end{equation}
Minimising the full potential we find two stationary points for real $X$
\begin{alignat}{1}
& X_0^\pm = \frac{1}{4 h^2 (\frac{1}{12} k \hat{\mu}^4 + m^2 \mu^2 \pi^2 \Gamma^2)} \bigg( -2 c h^3 k m \hat{\mu}^2 + 2 m \mu^2 \hat{\mu}^2 \pi^2 \Gamma^2 \pm \sqrt{2 \hat{\mu}^4}\nonumber \\
& \hspace{2cm} \times \sqrt{ (2 m^2 (c h^3 k - \mu^2 \pi^2 \Gamma )^2 - h^4 \mu^2 (N-k) (\frac{1}{12} k \hat{\mu}^4 + m^2 \mu^2 \pi^2 \Gamma^2 ) )} \; \bigg),
\end{alignat}
where we defined $c= \tilde{c} \mu^4 / \hat{\mu}^2$ and  $\Gamma = 1 + \gamma (N_f-k)$. One may derive simplifications for the case $\gamma=0$ by setting $\Gamma = 1$. We have two conditions for a stationary point to be a minimum 
\begin{eqnarray}
\left( \frac{X}{\mu} \right)^2 &>& \frac{\hat{\mu}^4 (N-k)}{8(\frac{1}{12} k \hat{\mu}^4 + m^2 \mu^2 \pi^2 \Gamma^2)}, \nonumber \\
0 &<& \hat{\mu}^4 \left( 2 m^2 (c h^3 k - \mu^2 \pi^2 \Gamma )^2 - h^4 \mu^2 (N-k) (\frac{1}{12} k \hat{\mu}^4 + m^2 \mu^2 \pi^2 \Gamma^2 ) \right) 
\end{eqnarray}
where the second one is the reality condition of the solutions. 
From this we may derive a lower bound on m which is set to be the smallest mass scale in this model. We get
\begin{equation} \label{bound_m}
m^2 > \frac{\frac{1}{12}h^4 k \mu^2 \hat{\mu}^4 (N-k)}{2 c^2 h^6 k^2 - 4 c h^3 k \mu^2 \pi^2 \Gamma + \mu^4 \pi^2 \Gamma^2 (2 \pi^2 - h^4 (N-k))}.
\end{equation}
The above is a generalization of a similar lower bound found in \cite{Giveon:2009yu}, (referred to as GKK from now on) but now includes the additional multitrace deformation
related to $\gamma$ and has also included a tree level $|X|^2$ term and a 1-loop linear term (proportional to $c$).÷\footnote{ In order to compare with the results in GKK the parameter $\epsilon$  defined by the latter is related to our mass scale 
$m$ via $m = \epsilon \hat{\mu}$ and  $\mu_i, i=1,2$ in the GKK notation corresponds to our $\mu, \hat{\mu}$. Finally the field $Z$ in GKK is our field $X$.}

As a result, we encounter a more complex expression for the lower bound in $m$ which requires us to also demand that the denominator of this bound 
on $m^2$ is positive. We may approximate the coefficient $c \sim 9/48 \mu^2 \Gamma + \dots$ since the term proportional to $\mu^2 \log [\mu^2]$ does not change the general analysis. Taking $h \sim 1$ we find 
\begin{equation}
9 k^2 + 224 k \pi^2 -128 \pi^2 (N- 2 \pi^2) > 0 
\end{equation}
in order to have a positive denominator in (\ref{bound_m}). 
For a given value of $N$, this constraint puts a lower bound on the allowed values of $k$. 

Using these approximations and the condition on $m^2$ we get 
\begin{equation}\label{xbound}
\left( \frac{X}{\mu} \right)^2 > \frac{3 (N-k)}{2 (k+J)}
\end{equation}
with
\begin{equation}
J = \frac{128 \pi^2  (h^4 k (N-k)) }{(9 k^2 + 224 k \pi^2 -128 \pi^2 (N- 2 \pi^2)) }.
\end{equation}
The lower bound in eq(\ref{xbound}) is an extension of the simpler expression, namely $(N-k)/k$, given by GKK. 
 It is interesting that this lower bound on the vev of $X$ is independent of  $\gamma$ (though the lower bound on $m$
does depend on both these parameters).   

Due to the hierarchy condition we must also demand that 
\begin{equation}
\frac{3 (N-k)}{2 (k+J)}<1.
\end{equation}
As such there will be constraints on the allowed values of $k$ for a given $N$ in which all bounds are satisfied
in the same spirit as those found in GKK. It is not difficult to find values of $N,k $ that satisfy this.

Let us discuss the scaling behaviour of the vev. Since we have $c \sim \mu^2$ we see that $m \gtrsim \hat{\mu}^2 / \mu$. From this we find that the vev of $X$ generally scales as
\begin{equation}
\langle X \rangle \sim \mu.
\end{equation}
Using the tree level part of the full scalar potential we may approximate the SUSY restored vacuum of this model and get
\begin{equation}
\langle X \rangle_{\mbox{\tiny  SUSY}} \sim \frac{\hat{\mu}^2}{m}.
\end{equation}
The difference between the two vacua of the theory is $\Delta X \sim \mu$. Hence, we find for the bounce action
\begin{equation}
S \sim \frac{(\Delta X)^4}{\Delta V} \sim \frac{\mu^4}{\hat{\mu}^4}
\end{equation}
which is parametrically large as required.  
The minimum of full rank arises when $\omega^{T}\omega=\hat{\mu}^{2}$.  
Again the bounce action scales as $S \sim \frac{(\Delta X)^4}{\Delta V} \sim \frac{\mu^4}{\hat{\mu}^4}$ and tunneling is suppressed.
The main results of GKK are therefore reproduced in our ${\rm SO}$ based ISS model even if we include the additonal meson deformation 
$\mbox{Tr}[\Phi^2]$ in the superpotential. 


\subsection{Gaugino and sfermion masses from the uplifted model}
The gaugino masses are
\begin{equation}
M_{\gl_r}=\frac{\alpha_r h^{2}m[\hat{\mu}^{2}+hm(-1+\gamma(k-N_{f})X_0)]}{4\pi[\mu^{2}-hmX_0]}.
\end{equation}
The sfermion masses contribution arises from 
\begin{equation}
\Lambda^{2}_{S}=\frac{h^2 [h^4 m^4\! + \!2 \mu^4\!- \!2 h^3 m^3 X_0 \!-\! 
    2 h m \mu^2 X_0\! + \!
    h^2 m^2 (4 \mu^2\! +\! {X_0}^2)] [F_{X_0}/h]^2}{2[\mu^{2}\!-\!hm X_0]^{2}[4\mu^{2}\!+\!(\!-hm\!+\!X_0)^2)]}.
\end{equation}
Finally the effective messenger number is 
\begin{equation}
N^{(\rho,Z)}_{eff}=\frac{2 h^2 m^2 [4 \mu^2 + (-h m + X_0)^2]}{h^4 m^4 + 2 \mu^4 - 2 h^3 m^3 X_0 - 2 h m \mu^2 X_0 + 
 h^2 m^2 [4 \mu^2 + {X_0}^2]}
\end{equation}

A novel feature of the uplifted model is that a new messenger is naturally introduced.  
The fundamental $\omega$ may be a messenger when embedding into $SO(N_{f}-k)_{f}$ and is the sole messenger if one attempted to
 embed into into $SO(N-k)_{c}$  !  The contributions to gaugino and sfermion masses can be easily determined from Appendix A:

\begin{equation}
M^{\omega}_{\gl_r}=\text{T}(\mb{r})R[\omega] (\frac{\alpha_r}{4 \pi})\frac{F_{X}}{\braket{X}}.
\end{equation}
\begin{equation}
m^{2}_{\tilde{f}}=2 \sum_{r} C^{r}_{\tilde{f}}(\frac{\alpha_r}{4 \pi})^{2}\text{T}(\mb{r}) R[\omega]\Lambda^{2}_{S}
\end{equation}
where 
\begin{equation}
\Lambda^{2}_{S}=\frac{|F_{X}|^{2}}{|\braket{X}|^{2}}.
\end{equation}
These are valid when the off diagonal terms are smaller than the diagonal terms of  the $X$ mass matrix.

The gaugino masses from the $X$ messenger are 
\begin{equation}
M^{X}_{\gl_r}=T[\mb{r}]R[X](\frac{\alpha_r}{4 h^{2}\pi^{2}})\frac{\braket{X} \tilde{f}_{\gamma}+\braket{X}^{2}\tilde{g}_{\gamma}]}{\sqrt{\tilde{b}\mu^{2} 
+m^{2}\pi^{2}(1+\gamma(N_{f}-k))^{2}}}
\end{equation}
the sfermion contribution is calculated using
\begin{equation}
\Lambda^{2}_{S,X}=(\frac{h^{4}}{\pi^{2}})\frac{[\braket{X} \tilde{f}_{\gamma}+\braket{X}^{2}\tilde{g}_{\gamma}]^{2}}{\tilde{b}\mu^{2} +m^{2}\pi^{2}(1+\gamma(N_{f}-k))^{2}}.
\end{equation}

\subsection{Beta function contributions to the visible sector}
 When we embed $SO(N_{f}-k)=SO(10)$, the additional contributions to beta functions are
\begin{equation}
\delta b_{2,3}(\mu_{R}<2h q_{0})= -12 \quad \delta b_{2,3}(2h q_{0}<\mu_{R}<\Lambda)= -12-(k+N) \quad  \delta b_{2,3}(\mu_{R}>\Lambda)=-N_{c}
\end{equation}
where $2hq_{0}$ is approximately the mass of the fundamental messenger fields.  For the embedding of $SO(k)_{f}=SO(10)$  The contributions to the beta functions are 
\begin{equation}
 \delta b_{2,3}(\mu_{R}<2h q_{0})=0 \quad \delta b_{2,3}(2h q_{0}<\mu_{R}<\Lambda)= -12-(k+N) \quad \delta b_{2,3}(\mu_{R}>\Lambda)=-N_{c}.
\end{equation}
We may also attempt a new embedding in which the remaining unhiggsed gauge symmetry is associated with $SO(10)$,  such that
$SO(N-k)_{c}=SO(10)$.  The contributions to the beta functions are 
\begin{equation}
 \delta b_{2,3}(\mu_{R}<2h q_{0})=0 \quad \delta b_{2,3}(2h q_{0}<\mu_{R}<\Lambda)= -N_{f} \quad \delta b_{2,3}(\mu_{R}>\Lambda)=-N_{f}.
\end{equation}
This new embedding is very different to the first two as in this case there is a new a single fundamental messenger $\omega$ which has a different representation to ($\rho,Z$) in the remaining unhiggsed gauge group $SO(N-k)_{c}$.
We would like $N>k=10$  and $N_{f}>k$ when embedding into  $SO(k)_{D}=SO(10)$. Taking $N=11$ as the minimum for $N$ we would need $N_{f}>27$ and $N_{c}>21$ to satisfy the UV and IR free conditions. The other condition is more flexible with an extra variable in $SO(N_{f}-k)$.


\section{Extending ISS?}

There is one specific challenge to general gauge mediation and specifically ISS models and that is the solution to the $\mu/B_{\mu}$
 problem (see for instance \cite{Komargodski:2008ax}).  ISS models build upon Seiberg duality, whereby each of the dual descriptions 
 have the same degrees of freedom.  Direct couplings between a hidden sector and the visible standard model would mean introducing 
 the standard model matter as singlets of the macroscopic hidden sector.  These extra degrees of freedom would generally 
 destroy the Seiberg duality of the original macroscopic and microscopic ISS descriptions.  Further motivation for direct couplings
  come from the exotic hidden sector matter; one would like decay channels into lighter visible sector fields.  If we accept that
   there must be direct couplings between the hidden sector and visible standard model fields then we are motivated to using dual
    pairs to describe supersymmetry breaking see for instance \cite{Berkooz:1997bb,Brodie:1996xm}.  These are Seiberg dual models 
	whereby there are two local gauge symmetries in each of the macroscopic and microscopic descriptions. One can then interpret 
	one of the local symmetries as hidden and the other one as visible. The result would be a Seiberg dual description of the whole 
	standard model GUT parent and the hidden sector. Much of the lessons of ISS models would carry through; SUSY breaking by rank 
	condition in the macroscopic picture and a SUSY restored minimum due to the non-perturbative superpotential, relatively light fields from 
	the pseudo-modulus and dynamical generation of messengers with gaugino mass contributions from R-symmetry breaking deformations. This is also a more natural setting in which to introduce 16plet spinors, which would aid in completing the unification. Of course the biggest challenge to this is that one of the duals in each pair is always a strongly coupled gauge group and would make the gaugino and sfermion contributions incalculable.  A parallel approach to duality and unification has been discussed in \cite{abel_direct_2008}.


\section{Conclusions}

This paper is the first investigation of SUSY particle phenomenology based on an $SO(N)$   $\mathcal{N}=1 $ SQCD ISS  model.  In the many models that have already been investigated in the literature,  based $\mathcal{N}=1$  $SU(N)$ SQCD ISS, it
was natural to consider weakly gauging a flavour group to $SU(5)$ as a standard model GUT. We have extended this to  $SO(10)$.  We studied two kinds of `minimal' $SO(10)$ embeddings by identifying the unbroken flavour symmetries with $SO(10)$, however one may consider larger global symmetries in which $SO(10)$ is just a subgroup.  The existence of a metastable susy breaking minima, the conditions for it to be long lived, as well as the basic features of the sparticle spectrum describing heavier sfermions and lighter gauginos are similar to those found in the $SU$ models studied.  One of the striking differences however, is the larger number of hidden sector multiplets charged with respect to the standard model. In terms of the $SU(5)$ decomposition of $SO(10)$, the pseudo-modulus field has an extra  $\bf{15}$ and  $\bar{\bf{15}}$  in addition to the $\bf{24}$ counterpart of the $SU(5)$ ISS models.  This presents us with a possible window to distinguish various ISS models in experiment and to ascertain the unifying GUT group at much lower energies than $M_{GUT}$.

There have been many suggestions in the literature to improve and make viable the basic `vanilla' ISS models based on $SU$ flavour groups.  In particular an R-symmetry prevents gaugino masses in pure ISS. This R-symmetry may be broken explicitly or spontaneously. In $SO$ models one is forced to consider explicit rather than spontaneous breaking of R-symmetry. We systematically investigated the deformations suggested in the $SU$ literature, that of baryonic deformations of magnetic quarks \cite{Abel:2008gv}, the KOO deformation \cite{Kitano:2006xg} and multitrace operators of magnetic quarks and the magnetic meson \cite{Essig:2008kz}.  We extended the multitrace meson deformation by looking at reducing the rank  $k < N $ of the  magnetic squark matrices of supersymmetry breaking. This can help make ISS models more viable by 
generating a vev for the $X$ meson field at tree level by destabilizing the origin; so called `uplifted vacuum'\cite{Giveon:2009yu}. We highlight that the particular suggestion of baryonic deformations of magnetic quarks by contraction of $SO(N)_{D}$ indices play no role in generating gaugino masses and is quite different than that of the $SU$ case explored in \cite{Abel:2008gv}. We also make clear that not only do the ISS models have the standard fundamental messengers, they also have messenger contributions from the $X$ pseudo-modulus field and its contributions are calculated due to the techniques outlined in Appendix A.

Our work can be extended in various ways. In particular one may want to extract a detailed picture of the phenomenology of $SO(10)$ ISS models such as have been carried out for a particular $SU$ deformation in \cite{Zur:2008zg}. One may look at embedding $SO(10)$ into a subgroup of a larger rank $SO$ flavour symmetry as has been done for the $SU(5)$ case \cite{Franco:2009wf}. A systematic study of symplectic type ISS models is also warranted.  Supergravity extensions of this model and at finite temperature have only been explored in the $SU(N)$ case \cite{Abel:2006cr, Abel:2006my, Fischler:2006xh, Craig:2006kx, Anguelova:2007at}. It has also been popular to discuss string and Brane constructions that realise the $SU(N)$ ISS setup \cite{Giveon:2007fk,Ooguri:2006bg,F,Argurio:2007qk,Argurio:2006ny,Franco:2006es}.
 
\vskip 1cm
  

\paragraph{Acknowledgements} 

We would like to thank Rodolfo Russo and Vincenzo Cal\`o for discussions and comments. MM would like to thank B. Keren-Zur for useful comments and the organisers and attendees of
the ``International workshop on supersymmetry and supersymmetry breaking'' at the IPPP Durham 2009. DK is supported by a Queen Mary University of London studentship.  MM is funded by STFC.

\newpage
\appendix


\section{SO(10) example of ``General Gauge Mediation''}



%


\begin{figure}
\centering \includegraphics{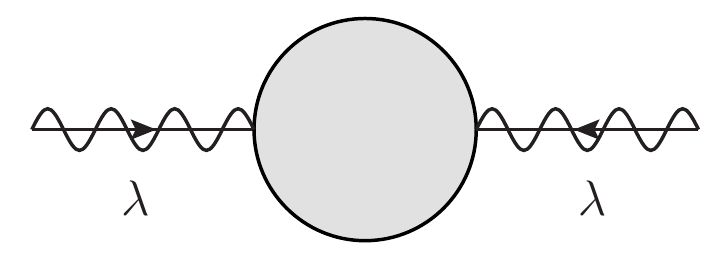}
\caption{Gaugino mass at one loop.}
\end{figure}

We give a brief review the techniques of  ``General Gauge Mediation'' \cite{Meade:2008wd}. In particular this section generalises, to adjoint representation, the results obtained for fundamental messengers in Appendix B of \cite{Marques:2009yu}. Further literature can be
 found in \cite{Buican:2008ws,Distler:2008bt,Ooguri:2008ez}. The gauge current superfield
\begin{equation}
\mathcal{J} \!= \!J +i\theta_{\alpha} j^{\alpha}+i\bar{\theta}^{\dot{\alpha}} j_{\dot{\alpha}}-
\theta_{\alpha}\sigma^{\mu \alpha\dot{\alpha}}\bar{\theta}_{\dot{\alpha}}j_{\mu}+
\frac{1}{2}\theta_{\alpha}\theta^{\alpha}\bar{\theta}^{\dot{\alpha}}\bar{\sigma}^{\mu}_{\dot{\alpha}\alpha}\partial_{\mu}j^{\alpha}-
\frac{1}{2}\bar{\theta}^{\dot{\alpha}}\bar{\theta}_{\dot{\alpha}}\theta^{\alpha}\bar{\sigma}^{\mu}_{\alpha\dot{\alpha}}\partial_{\mu}\bar{j}^{\dot{\alpha}}-
\frac{1}{4}\theta_{\alpha}\theta^{\alpha}\bar{\theta}^{\dot{\alpha}}\bar{\theta}_{\dot{\alpha}}\Box J
\end{equation}
is a real linear superfield defined by current conservation
\begin{equation}
\bar{D}^{2}\mathcal{J} =D^{2}\mathcal{J}=0.
\end{equation}
One can derive each component of the current multiplet by looking at the kinetic terms that couple the chiral superfields to
 the gauge vector supermultiplet.  Two parts are necessary to generate these currents, the vector super field in Wess Zumino gauge
\begin{equation}
V^{WZ}_{ij}=V^{a}T^{a}_{ij}=\theta\sigma^{\mu}\bar{\theta}A_{\mu}+\theta\theta\bar{\theta}\bar{\lambda}+\bar{\theta}\bar{\theta}\theta\lambda +
\frac{1}{2}\theta\theta\bar{\theta}\bar{\theta}D.
\end{equation}
and the full chiral multiplet
\begin{equation}
\Phi(x)=\phi(x)-i\theta\sigma^{\mu}\bar{\theta}\partial_{\mu}\phi(x)-\frac{1}{4}\theta^{2}\bar{\theta}^{2}\partial^{2}\phi(x)+
\sqrt{2}\theta\psi(x)+\frac{i}{\sqrt{2}}\theta^{2}\partial_{\mu}\psi(x)\sigma^{\mu}\bar{\theta}+\theta^{2}F(x).
\end{equation}
Once one has the gauge current supermultiplet, one may then take two point functions of the component currents.  
These two point functions are related to the soft terms for sfermions and gauginos.  For example to calculate the gaugino contribution one first looks at the fermionic current of the gauge current supermultiplet. In general this has the form
\begin{equation}
j^{A}_{\ga}(x)=-i\sqrt{2}(\phi^{*}T^{A}\psi(x)-\tilde{\phi}^{*}T^{A}\tilde{\psi}(x)).
\end{equation}
The scalar and fermionic fields in this three vertex source term are the interaction eigenstates of the messengers, whose external legs 
are gauginos $\gl$, which have been amputated, but whose generators $T^{A}$ are still part of the source. The gauge index $A$ runs from $1$ to the dimension of the gauge group.  The tilde represents the 
possibility of opposite charge conjugation of the field. The two point function is
\begin{equation}
\braket{j^{A}_{\alpha}(x)j^{B}_{\beta}(0)}=\epsilon_{\alpha\beta} \frac{T(\mb{r})\delta^{AB}}{x^{5}}B_{1/2}(x^{2}M^{2})
\end{equation}
where $M$ is a characteristic mass scale of the theory and $B_{1/2}$ is a complex function. $\text{T}(\mb{r})\delta^{AB}=\text{Tr}[T^{A}T^{B}]$, with $\text{T}(\mb{r})$ the Dinkin index of the messenger field. One can apply a Fourier transformation,
\begin{equation}
 M\tilde{B}_{1/2}(p^{2}/M)=\int d^{4}x e^{ip.x} \frac{1}{x^{5}}B_{1/2}(x^{2}M^{2})
\end{equation}
to obtain
\begin{equation}
\braket{j^{A}_{\alpha}(p)j^{B}_{\beta}(-p)}=\epsilon_{\alpha \beta}M T(\mb{r})\delta^{AB}\tilde{B}_{1/2}(p^{2}/M).
\end{equation}
The gaugino mass is given by
\begin{equation}
M_{\gl r}=g^{2}_{r}M T(\mb{r})\tilde{B}_{1/2}(0)
\end{equation}
$g_{r}$, with the gauge index $r$.  We have removed the $\delta^{AB}$ on both sides for clarity.  Ward identities are used to contract
 fields in the two point function where 
\begin{equation}
\braket{\phi(x)_{m}\phi(0)_{m}}=D(x,m)=\int \frac{d^{4}p}{(4\pi)^{4}}\frac{ie^{ip.x}}{p^{2}-m^{2}}.
\end{equation}
Similar expressions for the sfermion contributions may be found in \cite{Meade:2008wd}.

\begin{figure}
\centering
\includegraphics[width=15.5cm]{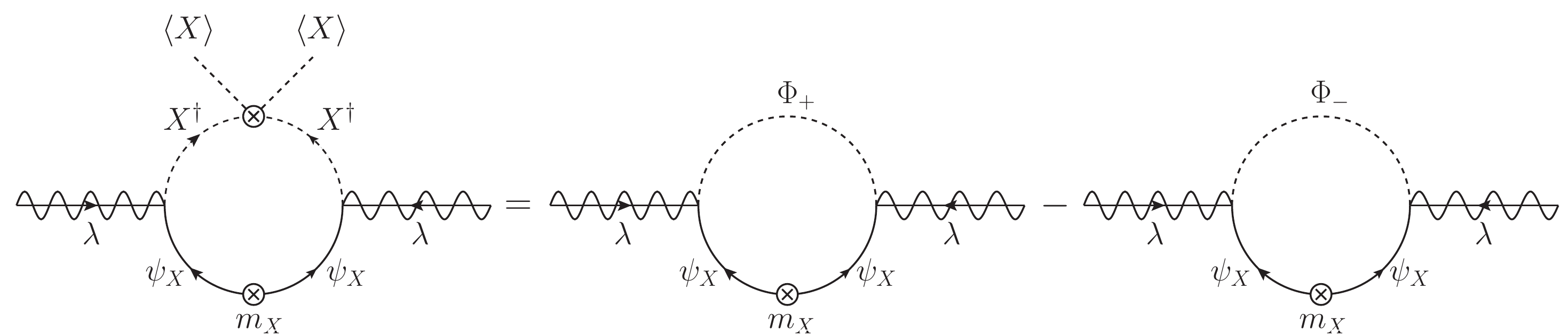}
\caption{One loop gaugino masses from the symmetric X messenger field.}
\end{figure}

Now we demonstrate the utility for the the case of $SO(10)$ for a symmetric representation field. 
These results would be similarly obtained in $SU(5)$ for the adjoint representation field and other high representations.
We start with the lagrangian
\begin{equation}
\mathcal{L}=\int d^{4}\theta (X^{\dagger}X+ X^{\dagger}\{V,X\})=\int d^{4}\theta( X^{\dagger}X+ 2X^{\dagger}_{ij}V_{jk}X_{ki}).
\end{equation}
The gauge field is in the antisymmetric (adjoint) and $ X $ is in the symmetric representation of $SO(10)$. We can amputate the gauge 
fields $(D^{A}, \lambda^{A}, \bar{\lambda}^{A}, A_{\mu}^{A})$ and leave the generators as part of the source currents. We will label 
the chiral multiplet as $X=(\phi_{X},\psi, F_{X})$. We obtain
\begin{alignat}{1}
J^{A}=&(\phi_{Xij}^{\dagger}T^{A}_{jk}\phi_{Xki}), \nonumber \\
j^{A}_{\alpha}=&-i\sqrt{2}(\phi_{Xij}^{\dagger}T^{A}_{jk}\psi_{ ki}), \nonumber \\
\bar{j}^{A}_{\dot{\alpha}}=&i\sqrt{2}(\bar{\psi}_{ij}T^{A}_{jk}\phi_{Xki}), \nonumber \\
j^{A}_{\mu}=&i(\partial_{\mu}\phi_{Xij}^{\dagger}T^{A}_{jk}\phi_{Xki}-\phi_{Xij}^{\dagger}T^{A}_{jk}\partial_{\mu}\phi_{Xki} )-\bar{\psi}_{ij} \sigma^{\mu}T^{A}_{jk}\psi_{ki}.
\end{alignat}
We add, to the kinetic Lagrangian, a mass matrix for the complex scalar X field:
\begin{equation}
M^{2}_{s}=
\left(
\phi_{X}^{\dagger} \phi_{X}
 \right) \left(
\begin{array}{cc}
a& b  \\
b & a
\end{array} \right) \left(
\begin{array}{c}
\phi_{X} \\
\phi_{X}^{\dagger}
\end{array} \right)
\end{equation}
We diagonalise the matrix to find the two real eigenvalues and their mass eigenstates,
\begin{equation}
\phi_{+}=\frac{1}{\sqrt{2}}(\phi_{X}+\phi^{\dagger}_{X})\quad i\phi_{-}=\frac{1}{\sqrt{2}}(\phi_{X}-\phi^{\dagger}_{X}).
\end{equation}
In terms of the interaction eigenstates these are,
\begin{equation}
\phi_{X}=\frac{1}{\sqrt{2}}(\phi_{+}+i\phi_{-})\quad \phi^{\dagger}_{X}=\frac{1}{\sqrt{2}}(\phi_{+}-i\phi_{-}).
\end{equation}
The masses are 
\begin{equation}
M^{2}_{\pm}=a\pm b \quad M_{\psi}=\sqrt{a}.
\end{equation}
We can rewrite the fermionic source which will couple to the gauginos, in terms of mass eigenstates
\begin{equation}
j^{A}_{\alpha}=-(\phi_{-}+i\phi_{+})T^{A}\psi_{X\alpha}
\end{equation}
Taking the two point function and using Ward identities we find,
\begin{equation}
\braket{j^{A}_{\alpha}j^{B}_{\beta}}= T(\mb{r})\delta^{AB}\braket{\psi_{X \alpha}\psi_{X \beta}}[\braket{\phi_{-}\phi_{-}}+
(i)^{2}\braket{\phi_{+}\phi_{+}}].
\end{equation}
In particular 
\begin{equation}
\braket{\psi_{X \alpha}(x)\psi_{X \beta}(0)}=\epsilon_{\alpha \beta} M_{\psi_{X}}D(x,M_{\psi_{X}}).
\end{equation}
We may substitute the propagators and evaluate,
\begin{alignat}{1}
M^{X}_{\gl_{r}}&=2M_{\psi_{X}}g^{2}_{r}R[X]T(\mb{r})(D(x;m_{X+})-D(x;m_{X-}))D(x;M_{\psi_{X}})\nonumber \\
&=2M_{\psi_{X}}g^{2}_{r}R[X]T(\mb{r})\int \frac{d^{4}k}{(2\pi)^{4}}(\frac{1}{k^{2}+m^{2}_{X+}}-\frac{1}{k^{2}+m^{2}_{X-}})\frac{1}{k^{2}+M^{2}_{\psi_{X}}}.
\label{sols2}
\end{alignat}
The resulting gaugino masses are 
\begin{equation}
M_{\gl_{r}}=T(\mb{r})R[X]\frac{\alpha_{r}}{4\pi}\frac{b}{\sqrt{a}}\times2g(x)
\end{equation}
where $x=b/a$ and 
\begin{equation}
g(x)=\frac{(1-x)\text{Log}(1-x)+(1+x)\text{Log}(1+x)}{x^{2}}.
\end{equation}
Similar reasoning results in an expression for the sfermions
\begin{alignat}{1}
m^{2}_{\tilde{f}}&=2\sum_{r}C^{r}_{\tilde{f}}(\frac{\alpha_{r}}{4\pi})^{2}\text{T}(\mb{r})R[X]\Lambda^{2}_{S}\\
\Lambda^{2}_{S}&=\frac{b^{2}}{a}f(x)
\end{alignat}
where 
\begin{equation}
f(x)=\frac{1+x}{x^{2}}[\ln(1+x)-2\text{Li}_{2}(x/[1+x])+\frac{1}{2}\text{Li}_{2}(2x/[1+x])]+(x\rightarrow-x).
\end{equation}

$\alpha_{r}$ is the gauge coupling at the messenger scale and $C^{r}_{\tilde{f}}$ denotes the quadratic Casimir of the irrep  $\tilde{f}$ of the gauge group labeled $r$. $R[X]$ is the rank of the representation of the $X$ field.

For the fundamental messengers the calculation is analogous. In the $SU(N)$ there is a  fundamental ($\phi_{i}$) and antifundamental ($\tilde{\phi}_{i}$)  messenger.  For $SO(N)$ the real and imaginary components of the fundamental messenger play these roles.  The source for fundamental messenger fields $(\rho,Z)$ are computed using 
\begin{alignat}{1}
J^{A}=&(\phi_{i}^{\dagger}T^{A}\phi_{i})\nonumber \\
j^{A}_{\alpha}=&-i\sqrt{2}(\phi_{i}^{\dagger}T^{A}\psi_{i})\nonumber \\
\bar{j}^{A}_{\dot{\alpha}}=&i\sqrt{2}(\bar{\psi}_{i}T^{A}\phi_{i})\nonumber \\
j^{A}_{\mu}=&i(\partial_{\mu}\phi_{i}^{\dagger}T^{A}_{i}\phi_{i}-\phi_{i}^{\dagger}T^{A}\partial_{\mu}\phi_{i} )-\bar{\psi}_{i} \sigma^{\mu}T^{A}\psi_{i}
\end{alignat}
The label $i$ is a flavour index.    The resulting gaugino and sfermions formulas are the same as the $SU(N)$ case \cite{Marques:2009yu}.

\newpage
\section{Viability of the second susy broken vacuum?}
Small mention has been made of the viability of the second susy broken vacuum in the literature.  We find it interesting to 
explore, as at least naively, the possibility of living in these new vacuum would remove (or at least reduce) the tension between 
the R-symmetry breaking terms need for gaugino masses and the metastability of the ISS vacuum.  In this discussion we start with a macroscopic 
$SU(N)\times SU(N)_{f}\times SU(N_{f}-N)_{f}$. The $SU(N)\times SU(N)_{f}$ global symmetry group becomes $SU(N)_{D}$ after 
use of the rank condition. We choose to embed in the $SU(N_{f}-N)_{f}$ global symmetry group. The field content is 
\begin{center}
\begin{tabular}{|ccc|}
\hline
Field & $SU(N)_{D} $& $SU(Nf-N)_{f} $\\
\hline
$\Phi = \left(
\begin{array}{cc}
Y_{\text{{\tiny $N$x$N$}}} & Z_{\text{{\tiny $N$x($N_{f}$-$N$)}}}  \\
\tilde{Z}_{\text{{\tiny ($N_{f}$-$N$)x$N$}}} & X_{\text{{\tiny ($N_{f}$-$N$)x($N_{f}$-$N$)}}}
\end{array} \right)_{\text{{\tiny $N_{f}$x$N_{f}$}}} $ &

$\left(
\begin{array}{cc}
\text{$\mathbf{Adj}+1$} & \bar{\text{\tiny\yng(1)}} \\
\text{\tiny\yng(1)}& 1
\end{array} \right) $ &

$\left(
\begin{array}{cc}
1 & \text{\tiny\yng(1)}  \\
\bar{\text{\tiny\yng(1)}} & \text{$\mathbf{Adj}+1$}
\end{array} \right)$\\

$\varphi=\left ( \begin{array}{c}
\chi_{N \times N} \\
\rho_{N_{f}-N\times N}
\end{array} \right)_{\text{\tiny{NfxN}}}$

&$\left (\begin{array}{c}
\text{$\mathbf{Adj}+1$}\\ \text{\tiny\yng(1)}\end{array} \right)$ 

& $\left(\begin{array}{c}
1 \\
 \bar{\text{\tiny\yng(1)}}
\end{array} \right)$\\

$\tilde{\varphi}=\left ( \begin{array}{c}
\tilde{\chi}_{N \times N} \\
\tilde{\rho}_{N\times N_{f}-N }
\end{array} \right)_{\text{\tiny{NxNf}}}$

&$\left (\begin{array}{c}
\text{$\mathbf{Adj}+1$}\\ \bar{\text{\tiny\yng(1)}}\end{array} \right)$ 

& $\left(\begin{array}{c}
1 \\
 \text{\tiny\yng(1)}
\end{array} \right)$ \\
\hline
\end{tabular}
\end{center} 

We will embed into the $SU(N_{f}-N)$.  We will make $N_{f}-N=6$ and $N=1$. With $N=N_{f}-N_{c}$ we find 
$N_{f}=7$ and $N_{c}=6$.  We use a canonical K\"ahler potential and the superpotential has an explicit R-symmetry breaking term:
\begin{equation}
W=h Tr[ \tilde{\chi}Y\chi +\tilde{\rho}X\rho +\tilde{\chi}Z\rho+ \tilde{\rho}\tilde{Z}\chi-\mu^2Y-\hat{\mu}^2X +h^2m_{z}\tilde{Z}Z].
\end{equation}
Generically there are two minima.  The ISS minima is parameterised by 
\begin{equation}
\tilde{\rho}\rho=0 \quad \tilde{\chi}\chi=\mu^2 I_{N} \quad \tilde{Z}=Z=0 \quad Y=0.
\end{equation}
 $X$ is a pseudo-modulus.  However the explicit R-symmetry breaking term generates $n$ new minima at 
\begin{alignat}{1}
&\chi^{s}_{\alpha}\chi^{s}_{\beta}=\mu^2 \mathbb{I}_{N} \nonumber\\
&\rho^{s}_{a} \rho^{s}_{b}=\frac{h^2 m^{2}_{z}}{\mu^2}ZZ=\text{diag}(\hat{\mu}^{2}...\hat{\mu}^{2},0...0)_{N_{f}-N}\nonumber\\
&X_{ab}=-\frac{\mu^2}{m_{z}}\text{diag}(1...1,\hat{x}...\hat{x})_{N_{f}-N}\\
&Y_{\alpha\beta}=-(\frac{\hat{\mu}^{2}}{m_{z}})\mathbb{I}_{N}\nonumber\\
&V_{low}=(N_{f}-N-n)|h\mu^{2}|^{2}\nonumber
\label{v}
\end{alignat}
where $n$ runs from $1$ to $N$. In this new minima the remaining $(N_{f}-N-n)$ components of $X_{ab}$ are pseudo-moduli, $\hat{x}$. 
 This is easy to see as, in the scalar potential, the $X$ field only appears in the term $V\supset |...+X\!\rho|^{2}$. 
 When $\rho$ is zero $X$ is necessarily a moduli. First we note that the $F_{X}$ terms of this new vacuum are
\begin{equation}
F_{X}=\text{Diag}(0,-h\hat{\mu}^2,...,-h\hat{\mu}^2)_{N_{f}-N}.
\end{equation}

Generalisations of the wavefunction renormalisation technique, in the regime that the $F$-term of the pseudo-modulus is
 smaller than the messenger scale, give analytic expressions for the gaugino and sfermion masses \cite{Cheung:2007es}. 
 For the gauginos  
\begin{equation}
m_{r}=\frac{\alpha_{r}}{4\pi}\Lambda_{G}=\frac{\alpha_{r}}{4\pi}F_{X}\sum_{i}\frac{\partial_{X}M_{i}}{M_{i}}
\end{equation}
where $M_{i}$ are the eigenvalues of the fermion mass matrix.  We now see why the embedding of $SU(5)$ into the whole of $SU(N_{f}-N)$ 
would have been an issue for this vacuum.  Also we cannot simply take out a tensor product of $I_{5}$ as the $X_{11}$ component gets a
 vev at tree level and the other $X_{ii}$ are pseudo-moduli.  Instead if we embed $SU(N_{f}-N)=SU(6)$, we can choose to rewrite the 
 original model in a form which makes explicit the field content under this new embedding group:
\begin{center}
\begin{tabular}{|ccc|}\hline
Field & $SU(N)_{D} $& $SU(Nf-2N)_{f} $\\
\hline
$\Phi \! = \! \left(
\begin{array}{ccc}
\! \! T_{\text{{\tiny $N$x$N$}}} & \tilde{S}_{\text{{\tiny $N$x$N$}}} & \tilde{M}_{\text{{\tiny $N$x($N_{f}$-$2N$)}}}  \! \! \\
\! \! S_{\text{{\tiny $N$x$N$}}} & Y_{\text{{\tiny $N$x$N$}}} & \tilde{Z}_{\text{{\tiny $N$x($N_{f}$-$2N$)}}} \! \! \\
\! \! M_{\text{{\tiny ($N_{f}$-$2N$)x$N$}}} & Z_{\text{{\tiny ($N_{f}$-$2N$)x$N$}}} & X_{\text{{\tiny ($N_{f}$-$2N$)x($N_{f}$-$2N$)}}} \! \! 
\end{array} \right)$ &

$\left(
\begin{array}{ccc}
\! \! \text{$\mathbf{Adj}\!+\!1$}  & \text{$\mathbf{Adj} \!+\! 1$} & \bar{\text{\tiny\yng(1)}}  \\
\! \! \text{$\mathbf{Adj}\!+\!1$} & \text{$\mathbf{Adj}\!+\!1$} & \bar{\text{\tiny\yng(1)}}   \\
\! \! \text{\tiny\yng(1)} & \text{\tiny\yng(1)} & 1
\end{array} \right) $ &

$\left(
\begin{array}{ccc}
1 &1& \text{\tiny\yng(1)}  \\
1 &1& \text{\tiny\yng(1)}  \\
\bar{\text{\tiny\yng(1)}} & \bar{\text{\tiny\yng(1)}} & \text{$\mathbf{Adj}\!+\!1$} 
\end{array} \right)$\\

$\varphi = \left ( \begin{array}{c}
\sigma_{\text{{\tiny $N $x$N$}}} \\
\chi_{\text{{\tiny $N $x $N$}}}\\
\rho_{\text{{\tiny ($N_{f}$-$2N$) x$N$ }}}
\end{array} \right)_{\text{\tiny{$N_{f}$x$N$}}}$ 

&$\left (\begin{array}{c}
\! \! \text{$\mathbf{Adj} + 1$} \! \! \\
\! \! \text{$\mathbf{Adj} + 1$} \! \! \\
\! \! \text{\tiny\yng(1)} \! \!
\end{array} \right)$ 

& $\left(\begin{array}{c}
1 \\
1 \\
 \bar{\text{\tiny\yng(1)}}
\end{array} \right)$\\

$\tilde{\varphi} = \left ( \begin{array}{c}
\tilde{\sigma}_{\text{{\tiny$ N$ x $N$}}} \\
\tilde{\chi}_{\text{{\tiny $N$ x $N$}}} \\
\tilde{\rho}_{\text{{\tiny  $N$x$N_{f}$-$2N$}} }
\end{array} \right)_{\text{\tiny{$N$x$N_{f}$}}}$

&$\left (\begin{array}{c}
\! \! \text{$\mathbf{Adj} + 1$} \! \! \\
\! \! \text{$\mathbf{Adj} + 1$} \! \! \\
\! \! \bar{\text{\tiny\yng(1)}} \! \! 
 \end{array} \right)$ 

& $\left(\begin{array}{c}
1 \\
1 \\
 \text{\tiny\yng(1)}
\end{array} \right)$
\\
\hline
\end{tabular}
\end{center}

As the second vacuum explicitly breaks the $SU(N_{f}-N)$ gauge symmetry to \newline $SU(N_{f}-N-n)$, we choose to gauge the $SU(N_{f}-N-n)$ to $SU(5)$.  This is viable.  One must however, add another meson deformation  $\delta W =h^{2}M_{m} \mbox{Tr} [\tilde{M}M ]$. We choose to parameterise the fermion mass matrix with the messengers ($\rho, Z, M$)
  and find
\begin{equation}
m_{f}= \mathbf{I}_{5}\otimes\left(
\begin{array}{ccc}
 h x   & h \mu & h\braket{\sigma} \\
 h \mu  &  h^2 m_{z} & 0 \\
 h\braket{\sigma} & 0 & h^{2}m_{m}  
\end{array}\right).
\end{equation}
The determinant of this matrix is non zero and gaugino masses can be generated.  This type of operator is natural from a multitrace meson deformation. This vacuum may be useful when exploring the types of model outlined in  \cite{Franco:2009wf}.


\begin{thebibliography}{10}

\bibitem{intriligator2006dsb}
K.~A. Intriligator, N.~Seiberg, and D.~Shih, ``{Dynamical SUSY breaking in
  meta-stable vacua},'' {\em JHEP} {\bf 04} (2006)  021,
\href{http://arxiv.org/abs/hep-th/0602239}{{\tt arXiv:hep-th/0602239}}.

\bibitem{Kitano:2006xg}
R.~Kitano, H.~Ooguri, and Y.~Ookouchi, ``{Direct mediation of meta-stable
  supersymmetry breaking},''
  \href{http://dx.doi.org/10.1103/PhysRevD.75.045022}{{\em Phys. Rev.} {\bf
  D75} (2007)  045022},
\href{http://arxiv.org/abs/hep-ph/0612139}{{\tt arXiv:hep-ph/0612139}}.

\bibitem{Abel:2007uq}
S.~A. Abel and V.~V. Khoze, ``{Metastable SUSY breaking within the standard
  model},''
\href{http://arxiv.org/abs/hep-ph/0701069}{{\tt arXiv:hep-ph/0701069}}.

\bibitem{Amariti:2006vk}
A.~Amariti, L.~Girardello, and A.~Mariotti, ``{Non-supersymmetric meta-stable
  vacua in SU(N) SQCD with adjoint matter},'' {\em JHEP} {\bf 12} (2006)  058,
\href{http://arxiv.org/abs/hep-th/0608063}{{\tt arXiv:hep-th/0608063}}.

\bibitem{abel_direct_2008}
S.~Abel and V.~V. Khoze, ``{Direct Mediation, Duality and Unification},''
  \href{http://dx.doi.org/10.1088/1126-6708/2008/11/024}{{\em JHEP} {\bf 11}
  (2008)  024},
\href{http://arxiv.org/abs/0809.5262}{{\tt arXiv:0809.5262 [hep-ph]}}.

\bibitem{Meade:2008wd}
P.~Meade, N.~Seiberg, and D.~Shih, ``{General Gauge Mediation},''
\href{http://arxiv.org/abs/0801.3278}{{\tt arXiv:0801.3278 [hep-ph]}}.

\bibitem{Seiberg:1994pq}
N.~Seiberg, ``{Electric - magnetic duality in supersymmetric nonAbelian gauge
  theories},'' \href{http://dx.doi.org/10.1016/0550-3213(94)00023-8}{{\em Nucl.
  Phys.} {\bf B435} (1995)  129--146},
\href{http://arxiv.org/abs/hep-th/9411149}{{\tt arXiv:hep-th/9411149}}.

\bibitem{Rajagopal:2000wf}
K.~Rajagopal and F.~Wilczek, ``{The condensed matter physics of QCD},''
\href{http://arxiv.org/abs/hep-ph/0011333}{{\tt arXiv:hep-ph/0011333}}.

\bibitem{Banks:2009cx}
T.~Banks and H.~E. Haber, ``{Note on the pseudo-Nambu-Goldstone Boson of
  Meta-stable SUSY Violation},''
\href{http://arxiv.org/abs/0908.2004}{{\tt arXiv:0908.2004 [hep-ph]}}.

\bibitem{Franco:2009wf}
S.~Franco and S.~Kachru, ``{Single-Sector Supersymmetry Breaking in
  Supersymmetric QCD},''
\href{http://arxiv.org/abs/0907.2689}{{\tt arXiv:0907.2689 [hep-th]}}.

\bibitem{Abel:2007jx}
S.~Abel, C.~Durnford, J.~Jaeckel, and V.~V. Khoze, ``{Dynamical breaking of
  $U(1)_{R}$ and supersymmetry in a metastable vacuum},''
  \href{http://dx.doi.org/10.1016/j.physletb.2008.01.065}{{\em Phys. Lett.}
  {\bf B661} (2008)  201--209},
\href{http://arxiv.org/abs/0707.2958}{{\tt arXiv:0707.2958 [hep-ph]}}.

\bibitem{Abel:2007nr}
S.~A. Abel, C.~Durnford, J.~Jaeckel, and V.~V. Khoze, ``{Patterns of Gauge
  Mediation in Metastable SUSY Breaking},''
  \href{http://dx.doi.org/10.1088/1126-6708/2008/02/074}{{\em JHEP} {\bf 02}
  (2008)  074},
\href{http://arxiv.org/abs/0712.1812}{{\tt arXiv:0712.1812 [hep-ph]}}.

\bibitem{Abel:2008gv}
S.~Abel, J.~Jaeckel, V.~V. Khoze, and L.~Matos, ``{On the Diversity of Gauge
  Mediation: Footprints of Dynamical SUSY Breaking},''
  \href{http://dx.doi.org/10.1088/1126-6708/2009/03/017}{{\em JHEP} {\bf 03}
  (2009)  017},
\href{http://arxiv.org/abs/0812.3119}{{\tt arXiv:0812.3119 [hep-ph]}}.

\bibitem{Martin:1997ns}
S.~P. Martin, ``{A Supersymmetry Primer},''
\href{http://arxiv.org/abs/hep-ph/9709356}{{\tt arXiv:hep-ph/9709356}}.

\bibitem{Dubovsky:1999xc}
S.~L. Dubovsky, D.~S. Gorbunov, and S.~V. Troitsky, ``{Gauge mechanism of
  mediation of supersymmetry breaking},'' {\em Phys. Usp.} {\bf 42} (1999)
  623--651,
\href{http://arxiv.org/abs/hep-ph/9905466}{{\tt arXiv:hep-ph/9905466}}.

\bibitem{Martin:1996zb}
S.~P. Martin, ``{Generalized messengers of supersymmetry breaking and the
  sparticle mass spectrum},''
  \href{http://dx.doi.org/10.1103/PhysRevD.55.3177}{{\em Phys. Rev.} {\bf D55}
  (1997)  3177--3187},
\href{http://arxiv.org/abs/hep-ph/9608224}{{\tt arXiv:hep-ph/9608224}}.

\bibitem{Giudice:1998bp}
G.~F. Giudice and R.~Rattazzi, ``{Theories with gauge-mediated supersymmetry
  breaking},'' \href{http://dx.doi.org/10.1016/S0370-1573(99)00042-3}{{\em
  Phys. Rept.} {\bf 322} (1999)  419--499},
\href{http://arxiv.org/abs/hep-ph/9801271}{{\tt arXiv:hep-ph/9801271}}.

\bibitem{Giudice:1997ni}
G.~F. Giudice and R.~Rattazzi, ``{Extracting Supersymmetry-Breaking Effects
  from Wave- Function Renormalization},''
  \href{http://dx.doi.org/10.1016/S0550-3213(97)00647-0}{{\em Nucl. Phys.} {\bf
  B511} (1998)  25--44},
\href{http://arxiv.org/abs/hep-ph/9706540}{{\tt arXiv:hep-ph/9706540}}.

\bibitem{Dimopoulos:1996ig}
S.~Dimopoulos and G.~F. Giudice, ``{Multi-messenger theories of gauge-mediated
  supersymmetry breaking},''
  \href{http://dx.doi.org/10.1016/S0370-2693(96)01513-4}{{\em Phys. Lett.} {\bf
  B393} (1997)  72--78},
\href{http://arxiv.org/abs/hep-ph/9609344}{{\tt arXiv:hep-ph/9609344}}.

\bibitem{Raby:2007yv}
S.~Raby, ``{SUSY Model Building},''
\href{http://arxiv.org/abs/0710.2891}{{\tt arXiv:0710.2891 [hep-ph]}}.

\bibitem{Amariti:2008uz}
A.~Amariti and A.~Mariotti, ``{Two Loop R-Symmetry Breaking},''
  \href{http://dx.doi.org/10.1088/1126-6708/2009/07/071}{{\em JHEP} {\bf 07}
  (2009)  071},
\href{http://arxiv.org/abs/0812.3633}{{\tt arXiv:0812.3633 [hep-th]}}.

\bibitem{Intriligator:2007py}
K.~A. Intriligator, N.~Seiberg, and D.~Shih, ``{Supersymmetry Breaking,
  R-Symmetry Breaking and Metastable Vacua},''
  \href{http://dx.doi.org/10.1088/1126-6708/2007/07/017}{{\em JHEP} {\bf 07}
  (2007)  017},
\href{http://arxiv.org/abs/hep-th/0703281}{{\tt arXiv:hep-th/0703281}}.

\bibitem{Barbieri:1982nz}
R.~Barbieri, S.~Ferrara, L.~Maiani, F.~Palumbo, and C.~A. Savoy, ``{QUARTIC
  MASS MATRIX AND RENORMALIZATION CONSTANTS IN SUPERSYMMETRIC YANG-MILLS
  THEORIES},''
\href{http://dx.doi.org/10.1016/0370-2693(82)90646-3}{{\em Phys. Lett.} {\bf
  B115} (1982)  212}.

\bibitem{Cheung:2007es}
C.~Cheung, A.~L. Fitzpatrick, and D.~Shih, ``{(Extra)Ordinary Gauge
  Mediation},'' \href{http://dx.doi.org/10.1088/1126-6708/2008/07/054}{{\em
  JHEP} {\bf 07} (2008)  054},
\href{http://arxiv.org/abs/0710.3585}{{\tt arXiv:0710.3585 [hep-ph]}}.

\bibitem{Abel:2009ze}
S.~A. Abel, J.~Jaeckel, and V.~V. Khoze, ``{Gaugino versus Sfermion Masses in
  Gauge Mediation},''
\href{http://arxiv.org/abs/0907.0658}{{\tt arXiv:0907.0658 [hep-ph]}}.

\bibitem{Zur:2008zg}
B.~K. Zur, L.~Mazzucato, and Y.~Oz, ``{Direct Mediation and a Visible
  Metastable Supersymmetry Breaking Sector},''
  \href{http://dx.doi.org/10.1088/1126-6708/2008/10/099}{{\em JHEP} {\bf 10}
  (2008)  099},
\href{http://arxiv.org/abs/0807.4543}{{\tt arXiv:0807.4543 [hep-ph]}}.

\bibitem{intriligator1995dmd}
K.~A. Intriligator and N.~Seiberg, ``{Duality, monopoles, dyons, confinement
  and oblique confinement in supersymmetric SO(N(c)) gauge theories},''
  \href{http://dx.doi.org/10.1016/0550-3213(95)00159-P}{{\em Nucl. Phys.} {\bf
  B444} (1995)  125--160},
\href{http://arxiv.org/abs/hep-th/9503179}{{\tt arXiv:hep-th/9503179}}.

\bibitem{Duncan:1992ai}
M.~J. Duncan and L.~G. Jensen, ``{Exact tunneling solutions in scalar field
  theory},''
\href{http://dx.doi.org/10.1016/0370-2693(92)90128-Q}{{\em Phys. Lett.} {\bf
  B291} (1992)  109--114}.

\bibitem{Jaeckel:2008jp}
J.~Jaeckel, ``{Metastable SUSY Breaking - Predicting the Fate of the
  Universe},'' \href{http://dx.doi.org/10.1016/j.nuclphysa.2009.01.022}{{\em
  Nucl. Phys.} {\bf A820} (2009)  83c--90c},
\href{http://arxiv.org/abs/0810.5307}{{\tt arXiv:0810.5307 [hep-ph]}}.

\bibitem{Slansky:1981yr}
R.~Slansky, ``{Group Theory for Unified Model Building},''
\href{http://dx.doi.org/10.1016/0370-1573(81)90092-2}{{\em Phys. Rept.} {\bf
  79} (1981)  1--128}.

\bibitem{Giveon:2007ef}
A.~Giveon and D.~Kutasov, ``{Stable and Metastable Vacua in SQCD},''
  \href{http://dx.doi.org/10.1016/j.nuclphysb.2007.11.037}{{\em Nucl. Phys.}
  {\bf B796} (2008)  25--43},
\href{http://arxiv.org/abs/0710.0894}{{\tt arXiv:0710.0894 [hep-th]}}.

\bibitem{Essig:2008kz}
R.~Essig, J.-F. Fortin, K.~Sinha, G.~Torroba, and M.~J. Strassler,
  ``{Metastable supersymmetry breaking and multitrace deformations of SQCD},''
  \href{http://dx.doi.org/10.1088/1126-6708/2009/03/043}{{\em JHEP} {\bf 03}
  (2009)  043},
\href{http://arxiv.org/abs/0812.3213}{{\tt arXiv:0812.3213 [hep-th]}}.

\bibitem{Xu:2007az}
F.-q. Xu and J.~M. Yang, ``{An Extension for Direct Gauge Mediation of
  Metastable Supersymmetry Breaking},''
\href{http://arxiv.org/abs/0712.4111}{{\tt arXiv:0712.4111 [hep-ph]}}.

\bibitem{Giveon:2009yu}
A.~Giveon, A.~Katz, and Z.~Komargodski, ``{Uplifted Metastable Vacua and Gauge
  Mediation in SQCD},''
\href{http://arxiv.org/abs/0905.3387}{{\tt arXiv:0905.3387 [hep-th]}}.

\bibitem{Komargodski:2008ax}
Z.~Komargodski and N.~Seiberg, ``{mu and General Gauge Mediation},''
  \href{http://dx.doi.org/10.1088/1126-6708/2009/03/072}{{\em JHEP} {\bf 03}
  (2009)  072},
\href{http://arxiv.org/abs/0812.3900}{{\tt arXiv:0812.3900 [hep-ph]}}.

\bibitem{Berkooz:1997bb}
M.~Berkooz, P.~L. Cho, P.~Kraus, and M.~J. Strassler, ``{Dual descriptions of
  SO(10) SUSY gauge theories with arbitrary numbers of spinors and vectors},''
  \href{http://dx.doi.org/10.1103/PhysRevD.56.7166}{{\em Phys. Rev.} {\bf D56}
  (1997)  7166--7182},
\href{http://arxiv.org/abs/hep-th/9705003}{{\tt arXiv:hep-th/9705003}}.

\bibitem{Brodie:1996xm}
J.~H. Brodie and M.~J. Strassler, ``{Patterns of duality in N = 1 SUSY gauge
  theories or: Seating preferences of theater-going non-Abelian dualities},''
  \href{http://dx.doi.org/10.1016/S0550-3213(98)00188-6}{{\em Nucl. Phys.} {\bf
  B524} (1998)  224--250},
\href{http://arxiv.org/abs/hep-th/9611197}{{\tt arXiv:hep-th/9611197}}.

\bibitem{Abel:2006cr}
S.~A. Abel, C.-S. Chu, J.~Jaeckel, and V.~V. Khoze, ``{SUSY breaking by a
  metastable ground state: Why the early universe preferred the
  non-supersymmetric vacuum},'' {\em JHEP} {\bf 01} (2007)  089,
\href{http://arxiv.org/abs/hep-th/0610334}{{\tt arXiv:hep-th/0610334}}.

\bibitem{Abel:2006my}
S.~A. Abel, J.~Jaeckel, and V.~V. Khoze, ``{Why the early universe preferred
  the non-supersymmetric vacuum. II},'' {\em JHEP} {\bf 01} (2007)  015,
\href{http://arxiv.org/abs/hep-th/0611130}{{\tt arXiv:hep-th/0611130}}.

\bibitem{Fischler:2006xh}
W.~Fischler, V.~Kaplunovsky, C.~Krishnan, L.~Mannelli, and M.~A.~C. Torres,
  ``{Meta-Stable Supersymmetry Breaking in a Cooling Universe},'' {\em JHEP}
  {\bf 03} (2007)  107,
\href{http://arxiv.org/abs/hep-th/0611018}{{\tt arXiv:hep-th/0611018}}.

\bibitem{Craig:2006kx}
N.~J. Craig, P.~J. Fox, and J.~G. Wacker, ``{Reheating metastable
  O'Raifeartaigh models},''
  \href{http://dx.doi.org/10.1103/PhysRevD.75.085006}{{\em Phys. Rev.} {\bf
  D75} (2007)  085006},
\href{http://arxiv.org/abs/hep-th/0611006}{{\tt arXiv:hep-th/0611006}}.

\bibitem{Anguelova:2007at}
L.~Anguelova, R.~Ricci, and S.~Thomas, ``{Metastable SUSY breaking and
  supergravity at finite temperature},''
  \href{http://dx.doi.org/10.1103/PhysRevD.77.025036}{{\em Phys. Rev.} {\bf
  D77} (2008)  025036},
\href{http://arxiv.org/abs/hep-th/0702168}{{\tt arXiv:hep-th/0702168}}.

\bibitem{Giveon:2007fk}
A.~Giveon and D.~Kutasov, ``{Gauge symmetry and supersymmetry breaking from
  intersecting branes},''
  \href{http://dx.doi.org/10.1016/j.nuclphysb.2007.04.033}{{\em Nucl. Phys.}
  {\bf B778} (2007)  129--158},
\href{http://arxiv.org/abs/hep-th/0703135}{{\tt arXiv:hep-th/0703135}}.

\bibitem{Ooguri:2006bg}
H.~Ooguri and Y.~Ookouchi, ``{Meta-stable supersymmetry breaking vacua on
  intersecting branes},''
  \href{http://dx.doi.org/10.1016/j.physletb.2006.08.035}{{\em Phys. Lett.}
  {\bf B641} (2006)  323--328},
\href{http://arxiv.org/abs/hep-th/0607183}{{\tt arXiv:hep-th/0607183}}.

\bibitem{F}
S.~Franco, I.~Garcia-Etxebarria, and A.~M. Uranga, ``{Non-supersymmetric
  meta-stable vacua from brane configurations},'' {\em JHEP} {\bf 01} (2007)
  085,
\href{http://arxiv.org/abs/hep-th/0607218}{{\tt arXiv:hep-th/0607218}}.

\bibitem{Argurio:2007qk}
R.~Argurio, M.~Bertolini, S.~Franco, and S.~Kachru, ``{Metastable vacua and
  D-branes at the conifold},'' {\em JHEP} {\bf 06} (2007)  017,
\href{http://arxiv.org/abs/hep-th/0703236}{{\tt arXiv:hep-th/0703236}}.

\bibitem{Argurio:2006ny}
R.~Argurio, M.~Bertolini, S.~Franco, and S.~Kachru, ``{Gauge/gravity duality
  and meta-stable dynamical supersymmetry breaking},'' {\em JHEP} {\bf 01}
  (2007)  083,
\href{http://arxiv.org/abs/hep-th/0610212}{{\tt arXiv:hep-th/0610212}}.

\bibitem{Franco:2006es}
S.~Franco and A.~M.~. Uranga, ``{Dynamical SUSY breaking at meta-stable minima
  from D- branes at obstructed geometries},'' {\em JHEP} {\bf 06} (2006)  031,
\href{http://arxiv.org/abs/hep-th/0604136}{{\tt arXiv:hep-th/0604136}}.

\bibitem{Marques:2009yu}
D.~Marques, ``{Generalized messenger sector for gauge mediation of
  supersymmetry breaking and the soft spectrum},''
  \href{http://dx.doi.org/10.1088/1126-6708/2009/03/038}{{\em JHEP} {\bf 03}
  (2009)  038},
\href{http://arxiv.org/abs/0901.1326}{{\tt arXiv:0901.1326 [hep-ph]}}.

\bibitem{Buican:2008ws}
M.~Buican, P.~Meade, N.~Seiberg, and D.~Shih, ``{Exploring General Gauge
  Mediation},'' \href{http://dx.doi.org/10.1088/1126-6708/2009/03/016}{{\em
  JHEP} {\bf 03} (2009)  016},
\href{http://arxiv.org/abs/0812.3668}{{\tt arXiv:0812.3668 [hep-ph]}}.

\bibitem{Distler:2008bt}
J.~Distler and D.~Robbins, ``{General F-Term Gauge Mediation},''
\href{http://arxiv.org/abs/0807.2006}{{\tt arXiv:0807.2006 [hep-ph]}}.

\bibitem{Ooguri:2008ez}
H.~Ooguri, Y.~Ookouchi, C.-S. Park, and J.~Song, ``{Current Correlators for
  General Gauge Mediation},''
  \href{http://dx.doi.org/10.1016/j.nuclphysb.2008.09.017}{{\em Nucl. Phys.}
  {\bf B808} (2009)  121--136},
\href{http://arxiv.org/abs/0806.4733}{{\tt arXiv:0806.4733 [hep-th]}}.

\end{thebibliography}

\providecommand{\href}[2]{#2}\begingroup\raggedright\endgroup

\end{document}